\documentclass[fleqn,usenatbib,useAMS]{mnras}

\usepackage[T1]{fontenc}
\usepackage{ae, aecompl}

\usepackage{graphicx}	% Including figure files
\usepackage{amsmath}	% Advanced maths commands
\usepackage{amssymb}	% Extra maths symbols
\usepackage{comment}
\usepackage[normalem]{ulem}       % for strikout \sout{} command
\PassOptionsToPackage{pdfpagelabels=false}{hyperref} 

\newcommand{\kapcor}{$w^{\kappa g}(\theta)$}
\newcommand{\gamcor}{$w^{\gamma_T g}(\theta)$}
\newcommand{\kapcormath}{w^{\kappa g}(\theta)}
\newcommand{\gamcormath}{w^{\gamma_T g}(\theta)}

\defcitealias{PlanckXIII:2015}{PlanckXIII}
\defcitealias{Jarvis:2015}{J15}

\title[Joint CMB and Galaxy Lensing]{Joint Measurement of
  Lensing-Galaxy Correlations Using SPT and DES SV Data}

\author[Baxter et al.]
{\parbox{\textwidth}{
E.~Baxter$^{1}$\thanks{E-Mail: ebax@sas.upenn.edu},
J.~Clampitt$^{1}$,
T.~Giannantonio$^{2,3,4}$,
S.~Dodelson$^{5,6}$,
B.~Jain$^{1}$,
D.~Huterer$^{7}$,
L.~Bleem$^{6,8,9}$,
T.~Crawford$^{6,10}$,
G.~Efstathiou$^{2,3}$,
P.~Fosalba$^{11}$,
D.~Kirk$^{12}$,
J.~Kwan$^{1}$,
C.~S{\'a}nchez$^{13}$,
K.~Story$^{14,15}$,
M.~A.~Troxel$^{16}$,
T. M. C.~Abbott$^{17}$,
F.~B.~Abdalla$^{12,18}$,
R.~Armstrong$^{19}$,
A.~Benoit-L{\'e}vy$^{12,20,21}$,
B.~Benson$^{5,6,10}$,
G.~M.~Bernstein$^{1}$,
R.~A.~Bernstein$^{22}$,
E.~Bertin$^{20,21}$,
D.~Brooks$^{12}$,
J.~Carlstrom$^{6,8,10}$,
A.~Carnero~Rosell$^{23,24}$,
M.~Carrasco~Kind$^{25,26}$,
J.~Carretero$^{11,13}$,
R.~Chown$^{27}$,
M.~Crocce$^{11}$,
C.~E.~Cunha$^{15}$,
L.~N.~da Costa$^{23,24}$,
S.~Desai$^{28,29}$,
H.~T.~Diehl$^{5}$,
J.~P.~Dietrich$^{28,29}$,
P.~Doel$^{12}$,
A.~E.~Evrard$^{7,30}$,
A.~Fausti Neto$^{23}$,
B.~Flaugher$^{5}$,
J.~Frieman$^{5,6}$,
D.~Gruen$^{15,31}$,
R.~A.~Gruendl$^{25,26}$,
G.~Gutierrez$^{5}$,
T.~de Haan$^{27,32}$,
G.~Holder$^{27}$,
K.~Honscheid$^{33,34}$,
Z.~Hou$^{6,8}$,
D.~J.~James$^{17}$,
K.~Kuehn$^{35}$,
N.~Kuropatkin$^{5}$,
M.~Lima$^{23,36}$,
M.~March$^{1}$,
J.~L.~Marshall$^{37}$,
P.~Martini$^{33,38}$,
P.~Melchior$^{19}$,
C.~J.~Miller$^{7,30}$,
R.~Miquel$^{13,39}$,
J.~J.~Mohr$^{28,29,40}$,
B.~Nord$^{5}$,
Y.~Omori$^{27}$,
A.~A.~Plazas$^{41}$,
C.~Reichardt$^{42}$,
A.~K.~Romer$^{43}$,
E.~S.~Rykoff$^{15,31}$,
E.~Sanchez$^{44}$,
I.~Sevilla-Noarbe$^{25,44}$,
E.~Sheldon$^{45}$,
R.~C.~Smith$^{17}$,
M.~Soares-Santos$^{5}$,
F.~Sobreira$^{5,23}$,
E.~Suchyta$^{1}$,
A.~Stark$^{46}$,
M.~E.~C.~Swanson$^{26}$,
G.~Tarle$^{7}$,
D.~Thomas$^{47}$,
A.~R.~Walker$^{17}$,
R.~H.~Wechsler$^{14,15,31}$
}
\vspace{0.4cm} \\
\parbox{\textwidth}{Author affiliations are listed at the end of this paper}}

\date{Last updated \today}

\pubyear{2016}

\begin{document}
\label{firstpage}
\pagerange{\pageref{firstpage}--\pageref{lastpage}}

\maketitle

% Abstract of the paper
\begin{abstract}
We measure the correlation of galaxy lensing and cosmic microwave
background lensing with a set of galaxies expected to trace the matter
density field.  The measurements are performed using pre-survey Dark
Energy Survey (DES) Science Verification optical imaging data and
millimeter-wave data from the 2500 square degree South Pole Telescope
Sunyaev-Zel'dovich (SPT-SZ) survey. The two lensing-galaxy
correlations are jointly fit to extract constraints on cosmological
parameters, constraints on the redshift distribution of the lens
galaxies, and constraints on the absolute shear calibration of DES
galaxy lensing measurements.  We show that an attractive feature of
these fits is that they are fairly insensitive to the clustering bias
of the galaxies used as matter tracers.  The measurement presented in
this work confirms that DES and SPT data are consistent with each
other and with the currently favored $\Lambda$CDM cosmological model.
It also demonstrates that joint lensing-galaxy correlation measurement
considered here contains a wealth of information that can be extracted
using current and future surveys.
\end{abstract}

% Select between one and six entries from the list of approved keywords.
% Don't make up new ones.

%%%%%%%%%%%%%%%%%%%%%%%%%%%%%%%%%%%%%%%%%%%%%%%%%%

%%%%%%%%%%%%%%%%% BODY OF PAPER %%%%%%%%%%%%%%%%%%

\section{Introduction}
\label{sec:intro}

Gravitational lensing of light from cosmological sources is sensitive
to both the matter content of the Universe and to its geometry
\citep[for a review see][]{Bartelmann:2010}.  A common approach to
measuring gravitational lensing with the goal of constraining
cosmology is to correlate some measure of the lensing strength with a
tracer of the matter density field, such as galaxies.  One advantage
of lensing-tracer cross-correlation measurements is that they
typically have much higher signal-to-noise than lensing-lensing
correlations.  Gravitational lensing of the cosmic microwave
background (CMB), for instance, was first detected \citep{Smith:2007}
by cross-correlating noisy CMB lensing maps with a catalog of radio
galaxies.  Similarly, in the context of galaxy lensing, the
cross-correlation of lensing induced shearing of background galaxies
with the positions of foreground galaxies (known as galaxy-galaxy
lensing) was detected \citep{Brainerd:1996} before shear-shear
correlations
\citep{Bacon:2000,Kaiser:2000,vanWaerbeke:2000,Wittman:2000}.
Furthermore, lensing-tracer cross-correlation measurements are
typically less sensitive to additive systematic errors since these
will tend to average to zero in the cross-correlation (assuming the
sources of systematic error are uncorrelated between the lensing
measurements and the tracer measurements, often a reasonable
approximation).  Henceforth, we will refer to sources of light used to
measure gravitational lensing distortion as {\it sources}, and we will
refer to objects used as tracers of the matter density field as {\it
  tracers}.

In this work, we perform a joint measurement of two lensing-tracer
cross-correlations that involve different sources, but the same set of
tracer galaxies.  The first source that we consider is the CMB, which
originates from a redshift of $z \sim 1100$. Gravitational lensing of
the CMB is typically measured in terms of the convergence, $\kappa$,
which quantifies the amount of lensing-induced dilation of an image
(defined rigorously in \S\ref{sec:formalism}).  We measure the angular
correlation, \kapcor{}, between the CMB-derived $\kappa$ and the
tracer galaxies.  The second source that we consider is a set of
galaxies at redshifts $0.8 \lesssim z \lesssim 1.3$.  Gravitational
lensing of galaxies is typically measured in terms of the shear,
$\gamma$, which quantifies the amount of lensing-induced stretching of
an image (defined rigorously in \S\ref{sec:formalism}).  We measure
the angular correlation, \gamcor{}, between the tracer galaxies and
the tangential shear, $\gamma_T$, which describes the component of the
shear perpendicular to the line connecting the image of a source
galaxy and a tracer galaxy.  The superscript $g$ in both \kapcor{} and
\gamcor{} is intended to remind the reader that these correlations are
with respect to the {\it same} set of tracer galaxies.

Because gravitational lensing is sensitive to the matter content of
the Universe and to its geometry, both \kapcor{} and \gamcor{} are
sensitive to cosmological parameters.  However, both
cross-correlations also depend on the {\it bias}, $b(k,z)$, of the
tracer galaxies, defined as the square root of the ratio of the tracer
power spectrum to the underlying matter power spectrum:
\begin{eqnarray}
b(k,z) \equiv \sqrt{P_{gg}(k,z) / P_{mm}(k,z),}
\end{eqnarray}
where $P_{gg}$ is the tracer galaxy power spectrum and $P_{mm}$ is the
matter power spectrum, both evaluated at wavenumber $k$ and redshift
$z$.  At large scales, the bias becomes scale-independent and is
therefore completely degenerate with the amplitude of the matter power
spectrum, $A_s$.  Other cosmological parameters, such as $\Omega_M$
(the matter density parameter) are also degenerate with the large
scale bias.  At small scales, the bias may become scale-dependent and
can therefore be degenerate with additional cosmological parameters.
For a single lensing-tracer cross-correlation measurement,
degeneracies between the tracer bias and the cosmological parameters
result in a degradation of cosmological constraints.

A joint measurement of two lensing-tracer cross-correlations that uses
different sources but the same set of tracer objects can break the
degeneracy between bias and cosmological parameters.  This basic idea
has been suggested by several authors. \citet{Jain:2003}, for
instance, propose measuring \gamcor{} with source galaxies in multiple
redshift bins at high redshift around a single set of tracer galaxies
at low redshift.  They show that in the limit that the tracer galaxies
are narrowly distributed in redshift, the ratio of the shear-tracer
correlation for one source bin to that of a different source bin is
completely insensitive to the bias of the tracer galaxies. Similarly,
\citet{Das:2009} propose measuring the ratio of two lensing-tracer
cross-power spectra, $C_{\ell}^{\kappa_{{\rm Gal}}
  g}/C_{\ell}^{\kappa_{{\rm CMB}} g}$, where $C_{\ell}^{\kappa_{{\rm
      Gal}} g}$ is the cross-power spectrum between a galaxy-lensing
derived $\kappa$ map and a set of tracer galaxies, $g$, and
$C_{\ell}^{\kappa_{{\rm CMB}} g}$ is the same quantity for a
CMB-lensing derived $\kappa$ map.  Again in the limit that the tracer
galaxies are narrowly distributed in redshift, the galaxy bias will
cancel in this ratio, making it a powerful cosmological probe.  An
attractive feature of combining CMB and galaxy lensing measurements
(compared to the galaxy-lensing-only measurement proposed by
\citealt{Jain:2003}) is that the large distance to the CMB last
scattering surface makes the lensing ratio more sensitive to
cosmological parameters \citep{Hu:2007}.

Note that unlike \citet{Das:2009}, we consider here a {\it joint fit}
to \kapcor{} and \gamcor{} rather than the ratio of two lensing-tracer
cross-correlations.  Performing a joint fit has several significant
advantages over the ratio measurement.  For one, the joint fit can be
applied directly to \kapcor{} and \gamcor{}.  This is advantageous
because CMB lensing is typically measured in terms of $\kappa$, while
galaxy lensing is typically measured in terms of $\gamma_T$, and the
conversion from $\gamma_T$ to $\kappa$ (or vice versa) is difficult
and potentially susceptible to systematic biases.  Furthermore, while
the probability distribution functions of the \kapcor{} and \gamcor{}
measurements can be reasonably approximated as multivariate Gaussians,
the ratio of two noisy Gaussian quantities is no longer Gaussian
distributed and is therefore difficult to model.  Additionally, when
the tracer galaxies do not have a very narrow redshift distribution,
the exact cancellation of the tracer bias in the ratio does not hold,
and the interpretation of the ratio measurement becomes difficult.
Finally, the joint fit contains more information than the ratio since
the ratio is computed from a combination of the two individual
\kapcor{} and \gamcor{} measurements.

In addition to being a powerful probe of cosmology, the joint
measurement of \kapcor{} and \gamcor{} can also be used to constrain
sources of systematic error that may impact either of the two measured
lensing-tracer correlations.  Sources of systematic error that affect
the CMB-derived $\kappa$ map are unlikely to have the same effect on
the galaxy-lensing derived $\gamma_T$ and vice versa.  Consequently,
joint measurement of both \kapcor{} and \gamcor{} can be used to
constrain systematic errors in the lensing measurements that would be
difficult (or impossible) to measure with a single lensing-tracer
cross-correlation \citep{Das:2013, Vallinotto:2013}.  Recently,
\citet{Liu:2016} used the joint measurement of galaxy and CMB lensing
around a set of tracer galaxies to constrain the multiplicative bias
of lensing shear measurements made by the Canada-France-Hawaii
Telescope Lensing Survey.  In addition to systematic errors in
the lensing measurements, the joint observable here is also sensitive
to systematic errors in the redshift distributions of the source and
tracer galaxies.  For DES (and other ongoing and future optical
surveys), the redshifts of the vast majority of galaxies are
determined photometrically.  Because photometric redshift estimation
is challenging, potentially subject to systematic errors, and requires
large spectroscopic training sets \citep[e.g.][]{Hildebrandt:2010},
the possibility of using the joint measurement of \kapcor{} and
\gamcor{} to constrain galaxy redshift distributions is appealing.

In this work, we measure \kapcor{} and \gamcor{} using data from the
South Pole Telescope (SPT) and pre-survey Dark Energy Survey (DES)
Science Verification (SV) imaging.  We perform a joint fit to the
measured \kapcor{} and \gamcor{} to extract constraints on
cosmological parameters, the photometric redshift distribution of the
tracer galaxies, and systematic biases in our measurements of
tangential shear (henceforth, we will refer to systematic biases as
{\it systematics} to eliminate confusion with the clustering bias).
Ultimately, DES will observe roughly 5000 sq. deg. of the Southern
sky; the SV data used in this work, however, is restricted to a small
fraction (roughly 3\%) of the full survey area.  Because of the small
area of the DES SV survey we do not expect to obtain highly
competitive constraints in this preliminary analysis.  For this
reason, we treat the measurement presented here mainly as a
consistency check between SPT and DES data and as a proof-of-principle
for the joint \kapcor{} and \gamcor{} measurement.

Our analysis builds upon other recently published analyses of DES SV
data.  The galaxy catalog used here was constructed and tested for
systematic effects in \citet{Crocce:2016} and references therein.  The
galaxy shear catalog used in this work was extensively tested in
\citet{Jarvis:2015}, while galaxy-galaxy lensing measurements and
systematics tests were performed in \citet{Clampitt:2016}.
Additionally, the cross-correlation between the galaxy catalog used in
this work and the SPT-derived CMB $\kappa$ map was first measured in
\citet{Giannantonio:2016}.  These and other DES SV papers provided key
methodological ingredients that support the analysis presented here.

In principle, one could imagine expanding the scope of the joint
\kapcor{} and \gamcor{} measurement considered here to include
additional correlations between galaxies, galaxy shear and CMB
$\kappa$.  In fact, all of the other possible correlations involving
these observables have already been measured by DES and SPT: the
shear-shear correlation was measured in \citet{DESshear2pt}, the
galaxy-galaxy correlation was measured in \citet{Crocce:2016}, and the
correlation between CMB $\kappa$ and galaxy shear was measured in
\citet{Kirk:2015}.  We have two reasons for restricting the analysis
in this work to \kapcor{} and \gamcor{}.  First, because \kapcor{} and
\gamcor{} are cross-correlations between different observable
types---namely galaxy positions and gravitational lensing
distortions---they are immune to several observational systematics.
Second, one of the main goals of this work is to show how degeneracies
between galaxy bias and parameters of interest are broken by
performing a joint measurement of galaxy and CMB lensing.  Since
neither the shear-shear correlation nor the correlation between galaxy
shear and CMB $\kappa$ depend on galaxy bias, their inclusion in this
analysis is not essential.  Since we are not attempting to generate
competitive cosmological constraints in this work, leaving out these
additional correlations is not a serious handicap for our analysis.

The outline of the paper is as follows.  In \S\ref{sec:formalism} we
introduce the necessary gravitational lensing formalism; in
\S\ref{sec:data} we describe the datasets used in this work; in
\S\ref{sec:measurement} we describe the process of measuring \kapcor{}
and \gamcor{}; in \S\ref{sec:analysis} we describe our models for the
data and the process of extracting constraints on the parameters of
these models.  Our results are presented in \S\ref{sec:results} and
conclusions are given in \S\ref{sec:discussion}.

\section{Formalism}
\label{sec:formalism}

We are interested in the cross-correlations of CMB lensing and galaxy
lensing with a single set of tracer galaxies.  We quantify CMB lensing
using the lensing convergence, $\kappa$, as there is a well developed
literature on estimating $\kappa$ from CMB temperature maps.  We
quantify galaxy lensing with the lensing shear, $\gamma$, as this
quantity can be measured directly from the distortion of galaxy
shapes.  In principle, one could convert $\kappa$ to $\gamma$ or vice
versa, but we do not take this approach here.

The convergence $\kappa$ is given by a weighted integral of the
distribution of matter along the line of sight.  Following the
notation of \citet{Jain:2003}, $\kappa$ in the direction specified by
the unit vector $\hat{n}$ is
\begin{eqnarray}
\kappa\left(\hat{n}\right) = \frac{3}{2c^2}\Omega_M H_0^2 \int d\chi \, g(\chi) \frac{\delta (\chi \hat{n}, \chi)}{a(\chi)},
\end{eqnarray}
where $\chi$ is the comoving distance, $a(\chi)$ is the scale factor,
$\delta(\chi \hat{n}, \chi)$ is the overdensity evaluated along the
line of sight, and we have assumed a spatially flat Universe.
Here $g(\chi)$ is the lensing weight function:
\begin{eqnarray}
\label{eq:g}
g(\chi) = \chi \int_{\chi}^{\infty} d\chi' \frac{(\chi' - \chi)}{\chi'} W^s (\chi'),
\end{eqnarray}
where $W^s(\chi)$ is the normalized distribution of the sources in
comoving distance.  The weight $W^s(\chi)$ is given in terms of the
distribution of sources as a function of redshift, $N^s(z)$, by
\begin{eqnarray}
\label{eq:norm_source_dist}
W^s(\chi) = \frac{1}{\int dz' N^s(z')} \frac{dz}{d\chi} N^s(z).
\end{eqnarray}

We define $n^g(\hat{n})$ as the projected density of tracer
galaxies in direction $\hat{n}$ and $\delta n^g(\hat{n}) = (
n^g(\hat{n}) - \bar{n}^g)/\bar{n}^g$.  The angular correlation
between $\kappa$ and the tracer galaxies is then $\kapcormath{} =
\langle \delta n^g(\hat{n}) \kappa(\hat{n}') \rangle$, where the
average is taken over all pairs of points chosen so that the angular
separation between $\hat{n}$ and $\hat{n}'$ is $\theta$.  As we
will discuss below, filtering of the CMB-derived $\kappa$ field in
harmonic space means that it is useful to express \kapcor{} in terms
of the cross-power spectrum, $ C_{\ell}^{\kappa g}$, between the
CMB-derived $\kappa$ and the tracer galaxies:
\begin{eqnarray}
\label{eq:wtheta}
\kapcormath{} = \sum_{l= 0}^{\infty} \left( \frac{2l + 1}{4\pi}\right) P_l(\cos \theta) C_{\ell}^{\kappa g},
\end{eqnarray}
where $P_l$ is the Legendre polynomial of order $l$.  This expression
is exactly correct on the curved sky.  At small angular scales, the
Limber approximation \citep{Limber:1953, Kaiser:1992} is valid and we
can relate $C_{\ell}^{\kappa g}$ to the matter power spectrum,
$P(k,\chi)$, through
\begin{multline}
C_{\ell}^{\kappa g} = \\
\frac{3\Omega_M H_0^2}{2c^2} \int d\chi
\frac{1}{\chi^2}\frac{g^{{\rm CMB}}(\chi) }{a(\chi)}W^g(\chi) b\left(
\frac{l}{\chi},\chi \right) P\left( \frac{l}{\chi},\chi \right),
  \label{eq:limber}
\end{multline}
where $H_0$ is the Hubble constant, $W^g(\chi)$ is the distribution of
tracer galaxies in comoving distance (defined analogously to
$W^s(\chi)$ for the sources), $b\left( k,\chi \right)$ is the
clustering bias of the tracer galaxies \citep{Bartelmann:2010,
  Das:2009}, and $g^{\rm CMB}(\chi)$ is the lensing weight function
for the CMB source. Here, we make the approximation that all of the
CMB light is sourced from a single comoving distance, $\chi_{*}$, so
that $W_s(\chi) = \delta(\chi - \chi^{*})$ and therefore $g^{{\rm
    CMB}}(\chi) = \left[ \chi (\chi^{*} - \chi)/\chi^{*} \right]
\Theta(\chi^{*} - \chi)$, where $\Theta(\chi)$ is the Heaviside step
function.  Our convention is that the forward Fourier transform is
defined by
\begin{eqnarray}
f(\vec{x}) = \int \frac{d^3 k}{(2\pi)^3}e^{i \vec{k} \cdot \vec{x}} \tilde{f}(\vec{k}),
\end{eqnarray}
and the power spectrum is related to $\delta$ by 
\begin{eqnarray}
\langle \delta(\vec{k}) \delta(\vec{k}') \rangle = (2\pi)^3 \delta^3 (\vec{k} - \vec{k}')P(k).
\end{eqnarray}

The angular correlation between the shears of the source galaxies and
the tracer galaxies is measured in terms of the tangential shear,
$\gamma_T$, the component of the shear orthogonal to the line
connecting the source galaxy at which the shear is measured to the
tracer galaxy:
\begin{eqnarray}
\gamma_T = -\gamma_{1} \cos(2\varphi) -\gamma_{2} \sin(2\varphi),
\end{eqnarray}
where $\gamma_{1}$ and $\gamma_{2}$ are the components of the shear,
$\gamma$, in a Cartesian basis, and $\varphi$ is the position angle of
the tracer galaxy relative to the source galaxy in the same Cartesian
basis.  Analogously to \kapcor{}, $\gamcormath{} = \langle \delta
n^g(\hat{n}) \gamma_T(\hat{n}') \rangle$, where again the average is
taken over all pairs of points such that $\hat{n}$ and $\hat{n}'$ have
an angular separation of $\theta$.  In this case, since we do not
apply any filtering to the measured shear field, we can directly
compute \gamcor{} using \citep{Jain:2003}
\begin{multline}
\gamcormath{} = \\
\frac{3 \Omega_M H_0^2}{4 \pi c^2}  \int d\chi \frac{g^s(\chi)}{a(\chi)}  W^g(\chi)
\int dk \, k b\left( \chi,k \right) P\left(\chi,k \right) J_{2} (k\chi\theta),
 \label{eq:limber_gamma}
\end{multline}
where $g^s(\chi)$ is the lensing weight for the source
galaxies computed using Eq.~\ref{eq:g} and
Eq.~\ref{eq:norm_source_dist}.  The $W^g$, $g^{{\rm CMB}}$ and $g^s$
factors that enter into the computation of \kapcor{} and \gamcor{} are
shown in Fig.~\ref{fig:redshift_factor_compare}.  Note that $W^g$ has
different units than $g^{{\rm CMB}}$ and $g^s$;
Fig.~\ref{fig:redshift_factor_compare} is only intended to show the
redshift ranges that contribute to these functions.  For this reason,
we have normalized all curves in
Fig.~\ref{fig:redshift_factor_compare} to the same maximum value.

Although this work is primarily concerned with the joint measurement
of \kapcor{} and \gamcor{}, it is instructive to consider the
information content of the ratio of these two observables.  For this
purpose, we can re-write \kapcor{} in a form more similar to
Eq.~\ref{eq:limber_gamma} (assuming no filtering is applied to the
CMB-derived $\kappa$ map).  We have \citep{Guzik:2001}
\begin{multline}
\kapcormath{} = \\
\frac{3 \Omega_M H_0^2}{4 \pi c^2}  \int d\chi \frac{g^{\rm CMB}(\chi)}{a(\chi)}  W^g(\chi)
\int dk \, k b\left( \chi,k \right) P\left(k, \chi \right) J_{0} (k\chi\theta).\\
 \label{eq:kap_eq2}
\end{multline}
Note that the only differences between Eq.~\ref{eq:kap_eq2} and
Eq.~\ref{eq:limber_gamma} are that the Bessel function of order two has
been replaced by a Bessel function of order zero, and $g^s$ has been
replaced by $g^{\rm CMB}$.  In the limit that the distribution of
tracer galaxies in comoving distance is very narrow, it can be
approximated with a Dirac $\delta$-function: $W(\chi) = \delta(\chi -
\chi^g)$. The ratio of the two observables then reduces to
\begin{eqnarray}
\frac{\kapcormath{}}{\gamcormath} \approx \frac{g^{{\rm CMB}}(\chi^g)}{g^s(\chi^g)} \frac{\int dk \, k  b(k,\chi^g) P(k,\chi^g) J_0(k \chi^g \theta) }{ \int dk \, k b(k,\chi^g) P(k,\chi^g) J_2(k \chi^g \theta) }. \label{eq:ratio}
\end{eqnarray}
In the limit that the bias is scale-independent (valid at large
scales), we have $b(k, \chi^g) = b(\chi^g)$ and the bias factors
in the numerator and denominator of Eq.~\ref{eq:ratio} will cancel.
This cancellation makes the ratio $\kapcormath{}/\gamcormath{}$
independent of the scale-independent bias.  This property is shared by
the lensing ratios of \citet{Jain:2003} and \citet{Das:2009} (although
in those cases, even the scale-dependent bias cancels in the lensing
ratio).

While the scale-independent bias cancels in the ratio of our two
observables, the factor $g^{{\rm CMB}}(\chi^g)/g^s(\chi^g)$ does not.
This quantity contains information about the distances to the tracer
galaxies, the source galaxies and the CMB; it therefore contains
information about cosmological parameters that affect the geometry and
expansion history of the Universe.  Furthermore, information about
systematics in either the CMB or galaxy-derived lensing measurements
is not expected to cancel in the ratio since such systematics are
likely uncorrelated between the CMB and galaxy lensing measurements.
Finally, information about the redshifts of the source and tracer
galaxies does not cancel in the ratio since this information is also
contained in the $g^{{\rm CMB}}(\chi^g)/g^s(\chi^g)$ factor.  Since
the information content of $\kapcormath{}/\gamcormath{}$ is preserved
in the joint \kapcor{} and \gamcor{} observable, we expect that our
analysis of the joint observable will yield constraints on cosmology,
systematics in the lensing measurements, and systematics in the galaxy
redshift distributions, even if we allow for significant freedom in
the tracer clustering bias.

\section{Data}
\label{sec:data}

The galaxy lensing measurements and the tracer galaxy catalog used in
this work are both derived from DES SV imaging data which has been
reduced from the raw survey data by the DES Data Management pipeline
\citep{Mohr:2012, Desai:2012}.  The CMB lensing data is derived from
CMB temperature maps generated from SPT observations
\citep{Carlstrom:2011}.  All of the data sets used in this work have
been discussed in recently published literature.  We therefore keep
the discussion of the data somewhat brief, and direct the reader to
the corresponding references for more detailed information.

\begin{figure}
  \includegraphics[width=\columnwidth]{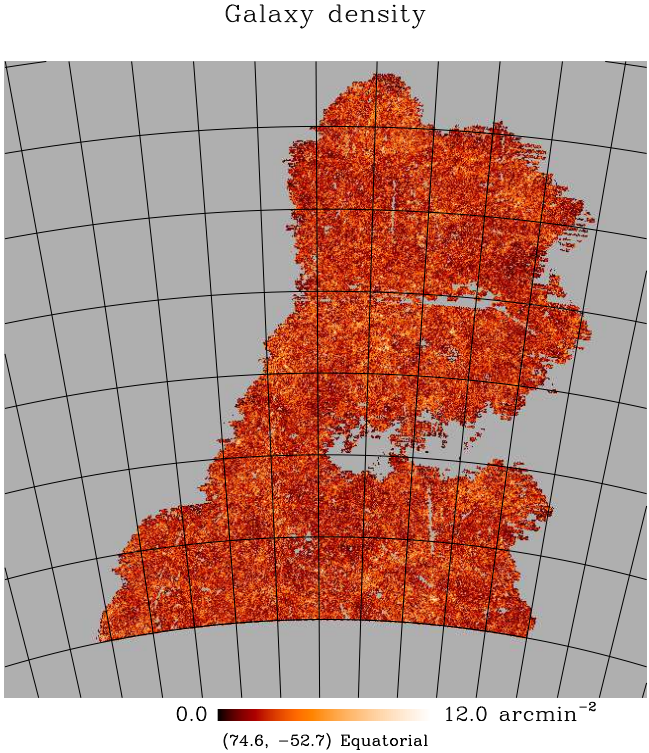}
  \caption{Density of tracer galaxies derived from the DES SV
    benchmark catalog plotted across the benchmark mask region.  The
    density map is shown at Healpix $N_{\rm side} = 2048$ resolution
    (corresponding to a pixel size of $\sim 1.7'$).  Note that
    although we plot the pixelized galaxy density here, \kapcor{} and
    \gamcor{} are computed using the un-pixelized tracer galaxy
    coordinates.  Grey regions are either masked or outside the SV
    footprint.  The grid lines are spaced 2.5 degrees apart in both
    R.A. and Dec.  The coordinates $(74.6, -52.7)$ indicate the
    position of the map center in R.A. and Dec.}
  \label{fig:tracer_map}
\end{figure}

\subsection{Data from the Dark Energy Survey}

\subsubsection{Tracer galaxy catalog}
\label{subsubsec:lens_catalog}

The catalog of galaxies used in this work as tracers of the matter
density field is derived from DES SV optical imaging data.  In total,
DES SV imaging covers roughly 300 sq. deg. of the southern sky that
was observed over 78 nights to near full-survey depth.  The analysis
here is restricted to the contiguous SPT-E field, which covers
approximately 139 sq. deg.  The DES SV final (`Gold') main galaxy
catalog\footnote{\texttt{http://des.ncsa.illinois.edu/releases/sva1}}
contains 25,227,559 galaxies.

The tracer galaxy catalog that we use in this work is a subset of the
full `Gold' catalog that was selected by \citet{Crocce:2016}, and
which is termed the {\it benchmark} selection.  Briefly, the benchmark
selection restricts the galaxy sample to $18 < i < 22.5$, where $i$ is
the \verb!MAG_AUTO!  quantity output by SExtractor
\citep{Bertin:1996}. Several additional cuts are applied that, for
instance, remove outliers in color space and remove stars that may be
falsely classified as galaxies.  The end result is a flux-limited
sample of galaxies over an area of roughly 131 sq. deg.  We use the
corresponding benchmark galaxy angular mask in this analysis.

%Note that the benchmark galaxy selection employed here is a slightly
%different version of that employed in \citet{Giannantonio:2016},
%reflecting minor updates to the benchmark catalog.

We use photometric redshift (photo-$z$) estimates for the purposes of
selecting tracer and source galaxies, and also for computing the
distributions of these two populations as a function of redshift
(necessary when we model the measured lensing-tracer
cross-correlations).  The photo-$z$ estimates used here are generated
using the neural network-based \texttt{skynet2} code
\citep{Graff:2014, Bonnett:2015}.  \texttt{skynet2} computes the
redshift probability distribution functions, $p(z)$, for each galaxy,
given the photometric colors of that galaxy.  Several photometric
redshift estimation codes have been applied to DES SV galaxies.  In
this work we use the \texttt{skynet2} code as it performed the best in
tests \citep{Bonnett:2015} and because this matches the choice made
for the cosmic shear analysis of DES SV data by \citet{DESshear2pt}.
\citet{Bonnett:2015} showed that \texttt{skynet2} was able to recover
the mean redshift of samples of DES SV `Gold' galaxies to typically
better than $0.04$.  In general, though, DES SV science results have
been shown to be quite robust to the choice of photo-$z$ estimation
code \citep[e.g.][]{Giannantonio:2016, Crocce:2016}.  Tracer and
source galaxies are selected on the basis of the $z$ value at which
$p(z)$ peaks, $z^p$.  For the tracers, we restrict the analysis to
galaxies with $0.4 < z^p < 0.8$.  The final tracer catalog contains
approximately 1.3 million galaxies.  A map of the tracer galaxy
density across the benchmark mask is shown in
Fig.~\ref{fig:tracer_map}.

\begin{figure}
  \includegraphics[width=\columnwidth]{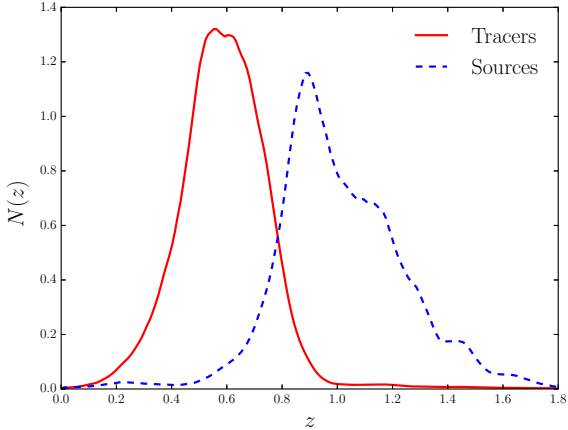}
  \caption{The normalized photometric redshift distributions, $N(z)$,
    for the tracer and source galaxy samples.  The tracers are selected
    using a $0.4 < z^p < 0.8$ cut, where $z^p$ is the redshift that
    maximizes the photometrically-determined redshift probability
    distribution for an individual galaxy, $p(z)$.  The sources are
    selected using a $0.8 < z^p < 1.3$ cut.}
  \label{fig:dndz}
\end{figure}

\begin{figure}
  \includegraphics[width=\columnwidth]{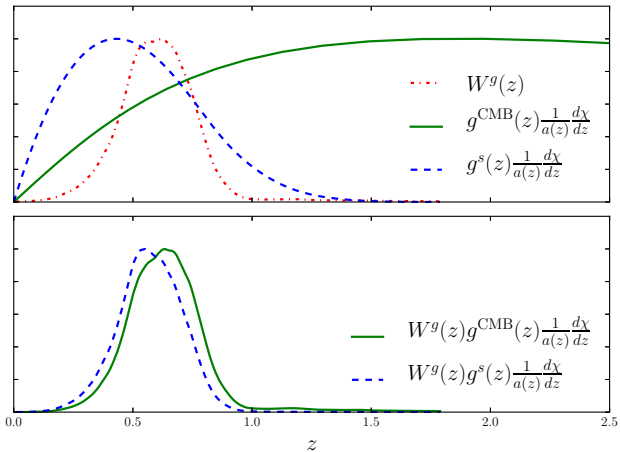}
  \caption{The $W^{g}$, $g^{{\rm CMB}}$ and $g^{s}$ factors (and the
    relevant products of these factors) that enter into the
    computation of \kapcor{} (Eqs.~\ref{eq:limber} and
    \ref{eq:wtheta}) and \gamcor{} (Eq.~\ref{eq:limber_gamma}).  The
    figure is intended to illustrate the redshift ranges that
    contribute most to \kapcor{} and \gamcor{}.  All curves have been
    normalized to the same maximum value. }
  \label{fig:redshift_factor_compare}
\end{figure}

The normalized $N(z)$ for the entire tracer catalog (i.e. the sum of
all the individual $p(z)$) is shown in Fig.~\ref{fig:dndz}.  The
corresponding $W^g(\chi)$ is shown in
Fig.~\ref{fig:redshift_factor_compare}, along with $g^s(\chi)$ and
$g^{{\rm CMB}}(\chi)$ for comparison (note that we have transformed
these quantities into functions of redshift for plotting purposes).
It is clear from Fig.~\ref{fig:redshift_factor_compare} that the
tracer galaxy $W^g(\chi)$ peaks in a redshift range for which both
$g^s(\chi)$ and $g^{{\rm CMB}}(\chi)$ have significant support.
%
%Using higher
%redshift tracer galaxies would increase the amplitude of \kapcor{},
%but would decrease the amplitude of \gamcor{}.  Similarly, using
%somewhat lower redshift tracer galaxies would increase the amplitude
%of \gamcor{}, but would decrease the amplitude of \kapcor{}. 
%
We note here that the measured $N(z)$ for the tracer catalog enters
into the modeling of \kapcor{} and \gamcor{} through $W^g(\chi)$; as
we will discuss more in \S\ref{subsubsec:redshift_bias_param}, the
dependence of \kapcor{} and \gamcor{} on $N(z)$ makes the joint
measurement of these quantities a potentially powerful probe of galaxy
redshift distributions.

\subsubsection{Source galaxy shear catalog}
\label{subsubsec:gal_shears}

The shear catalog used in this work to measure \gamcor{} is also
derived from DES SV data\footnote{The shear catalog is available at
  \texttt{http://des.ncsa.illinois.edu/releases/sva1}.}.  Two shear
catalogs were produced and tested extensively in \citet{Jarvis:2015}
(hereafter \citetalias{Jarvis:2015}): the
\texttt{ngmix}\footnote{\texttt{https://github.com/esheldon/ngmix}}
\citep{Sheldon:2014} and the
\texttt{im3shape}\footnote{\texttt{https://bitbucket.org/joezuntz/im3shape}}
\citep{Zuntz:2013} catalogs. We use only the \texttt{ngmix} catalog in
this work because they have a higher source number density.
%The
%exponential disk model option of \texttt{ngmix} and an expectation
%maximization \citep{Dempster:1977} method for fitting the point spread
%function.  
Shear estimation with \texttt{ngmix} was carried out using images in
$r, i, z$ bands. See \citetalias{Jarvis:2015} for more details and various
tests of the shear pipeline.  These choices are consistent with other
analyses of DES SV data, including the cosmology analysis of the
cosmic shear two-point function \citep{DESshear2pt}. \citetalias{Jarvis:2015}
performed many comparisons of the two shear pipelines, finding
generally good agreement.

Particularly relevant for our purposes is the \citetalias{Jarvis:2015}
comparison of the \texttt{im3shape} and \texttt{ngmix} tangential
shear measurements.  \citetalias{Jarvis:2015} measured tangential
shears around luminous red galaxies using both pipelines over an
angular range similar to that used here.  \citetalias{Jarvis:2015}
found that the ratio of the \texttt{im3shape} to \texttt{ngmix}
tangential shear measurements is consistent with expectations from the
application of these two shear pipelines to simulated data. The two
pipelines can therefore be considered consistent with each other in
their measurements of tangential shear.  Note, though, that this ratio
test does not preclude the possibility that both shear catalogs are
biased by a similar multiplicative factor; we will consider how the
joint measurement of \kapcor{} and \gamcor{} can be used to constrain
such multiplicative biases in \S\ref{subsubsec:shear_bias_param}.
  
We restrict the source catalog to galaxies with $0.8 < z^p < 1.3$.
This redshift cut and the various benchmark selections yield $\sim
947,000$ total source galaxies with a number density of $1.9/{\rm
  arcmin}^2$. The photometrically-determined $N(z)$ for the source
galaxies is shown in Fig.~\ref{fig:dndz}.

\subsection{Data from the South Pole Telescope}
\label{subsec:kappa_map}

\begin{figure}
  \includegraphics[width=\columnwidth]{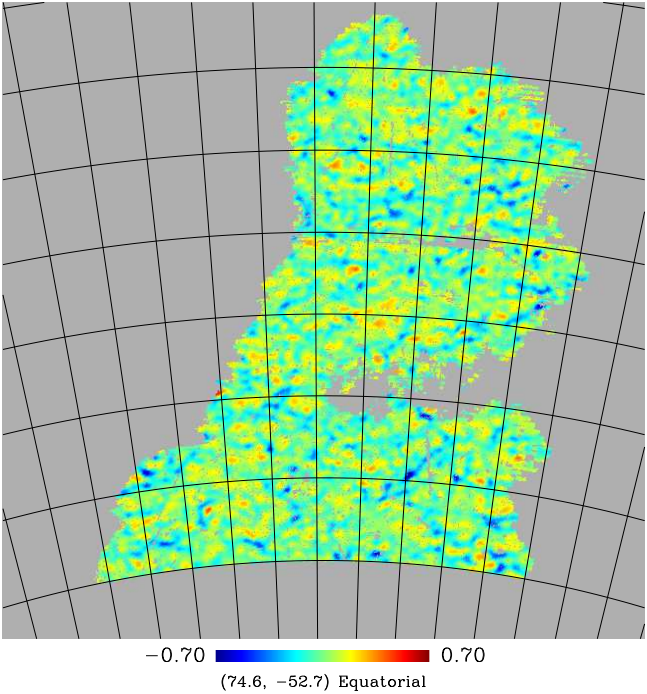}
  \caption{The filtered lensing convergence, $\kappa$, derived from
    SPT CMB data across the benchmark mask region.  As described in
    the text, the $\kappa$ map is high-pass filtered to $\ell > 30$
    and is smoothed with a Gaussian beam with $\theta_{\rm FWHM} =
    5.4'$.  For this plot we have also applied a Gaussian beam with
    $\theta_{\rm FWHM} = 10'$ to improve the visualization.  The map
    is shown at Healpix $N_{\rm side} = 2048$ resolution.  Note that
    although we have applied the benchmark mask in making this plot,
    the full SPT-derived $\kappa$ map is used when measuring
    \kapcor{}.  Coordinate system and gridlines are the same as those
    in Fig.~\ref{fig:tracer_map}.}
  \label{fig:kappa_map}
\end{figure}

The CMB $\kappa$ maps used in this work were derived from CMB
temperature data taken as part of the 2500 square degree South Pole
Telescope Sunyaev-Zel'dovich (SPT-SZ) survey \citep{Story:2013}.  Many
observations at 150 GHz of the SPT-E region were combined using
inverse-variance weighting to generate a $25^{\circ} \times
25^{\circ}$ CMB temperature map.  A CMB $\kappa$ map was then derived
from the CMB temperature map following the methods outlined in
\citet{vanEngelen:2012}, which rely on the quadratic estimator of
\citet{Hu:2001} and \citet{Hu:2002}.  The CMB $\kappa$ map was
pixelized using a Healpix \citep{Healpix:2005} grid with $N_{{\rm
    side}} = 2048$.  The processed SPT CMB lensing maps used here are
identical to those used in \citet{Giannantonio:2016} and we refer the
reader to that work for more details.  The same maps were also used in
the cross-correlation of CMB lensing with galaxy lensing measurement
of \citet{Kirk:2015}.  As in \citet{Giannantonio:2016}, we filter the
pixelized CMB $\kappa$ map to remove modes with $\ell < 30$ and also
apply Gaussian-beam smoothing with $\theta_{{\rm FWHM}} = 5.4'$.  In
our analysis, we use the CMB $\kappa$ map across the full SPT-E region
without applying any additional masking.  This means that the tracer
galaxies are correlated with regions of the CMB $\kappa$ map that lie
outside of the benchmark mask discussed
\S\ref{subsubsec:lens_catalog}.  The CMB $\kappa$ map is plotted in
Fig.~\ref{fig:kappa_map}.  To aid with visualization and comparison to
Fig.~\ref{fig:tracer_map} we have applied additional smoothing to the
$\kappa$ map in Fig.~\ref{fig:kappa_map} and have restricted the plot
to the benchmark mask.

\citet{Planck2015:XV} have also released a CMB-lensing-derived
$\kappa$ map that could be used to measure \kapcor{}.  As demonstrated
in \citet{Giannantonio:2016}, the signal-to-noise of \kapcor{}
measured using the benchmark galaxies and the Planck $\kappa$ map is
only slightly lower than the signal-to-noise of the same measurement
using the SPT $\kappa$ map.  However, because this work is intended as
a ``proof of concept'' for the joint \kapcor{} and \gamcor{}
measurement, we postpone a joint measurement of \kapcor{} and
\gamcor{} with Planck and DES data to future work based on a larger
DES sample.

\section{\texorpdfstring{$w^{\kappa g}(\theta)$}{w} and \texorpdfstring{$w^{\gamma_T g}(\theta)$}{w} Measurements}
\label{sec:measurement}

We measure \kapcor{} using the CMB $\kappa$ map described in
\S\ref{subsec:kappa_map} and the galaxy tracer catalog described in
\S\ref{subsubsec:lens_catalog}.  We estimate \kapcor{} with
\begin{eqnarray}
\label{eq:kap_estimator}
 \hat{w}^{\kappa g} (\theta_{\alpha}) &=& \bar{\kappa}_{\alpha} -  \bar{\kappa}^{{\rm rand}}_{\alpha},
\end{eqnarray}
where $\theta_{\alpha}$ is the (logarithmic) center of the $\alpha$
angular bin and
\begin{eqnarray}
\label{eq:kappa_sum}
 \bar{\kappa}_{\alpha} = \frac{\sum_i^{N_{{\rm pix}}} \sum_j^{N_{\rm tracer}} \kappa_j f_{ij\alpha}}{ \sum_i^{N_{{\rm pix}}} \sum_j^{N_{\rm tracer}}f_{ij\alpha}}.
\end{eqnarray}
Here, $N_{\rm pix}$ is the number of pixels in the $\kappa$ map,
$N_{\rm tracer}$ is the number of tracer galaxies, and $\kappa_j$ is
the value of $\kappa$ in the $j$th pixel. The weight function
$f_{ij\alpha} = 1$ if the angular separation between tracer galaxy $i$
and pixel $j$ is in angular bin $\alpha$ and $f_{ij\alpha} = 0$
otherwise.  The quantity $\bar{\kappa}_{\alpha}^{{\rm rand}}$ in
Eq.~\ref{eq:kap_estimator} is defined similarly to
$\bar{\kappa}_{\alpha}$, except that the tracer catalog is replaced by
a catalog of randomly distributed points that have the same angular
mask as the tracer galaxies.  By subtracting $\bar{\kappa}^{{\rm
    rand}}_{\alpha}$ from $\bar{\kappa}_{\alpha}$ we correct for any
mask and pixelization effects.  The sums in Eq.~\ref{eq:kappa_sum} are
computed using the publicly available tree code
\texttt{TreeCorr}\footnote{\texttt{https://github.com/rmjarvis/TreeCorr}}
\citep{Jarvis:2004}.

We measure \gamcor{} using the shear catalog described in
\S\ref{subsubsec:gal_shears} and the tracer galaxy catalog described
in \S\ref{subsubsec:lens_catalog}.  The estimator for \gamcor{} is
similar to that for \kapcor{}:
\begin{eqnarray} \label{eq:gammat-est}
\hat{w}^{\gamma_T g}(\theta_{\alpha}) &=& \bar{\gamma}_{\alpha} - \bar{\gamma}_{\alpha}^{\rm random},
\end{eqnarray}
where the $\gamma$ are understood to be tangential shears (dropping
the subscript $T$ temporarily for notational convenience), and
\begin{eqnarray}
 \bar{\gamma}_{\alpha} = \frac{\sum_i^{N_{{\rm source}}} \sum_j^{N_{\rm tracer}} \gamma_{ij} f'_{ij\alpha}}{ \sum_i^{N_{{\rm source}}} \sum_j^{N_{\rm tracer}}f'_{ij\alpha}}.
\end{eqnarray}
The sum over $i$ now runs over all source galaxies (instead of pixels)
and $\gamma_{ij}$ is the tangential shear of source galaxy $i$
measured with respect to tracer galaxy $j$.  The weight function
$f'_{ij\alpha}$ is defined such that $f'_{ij\alpha} = 1/(\sigma_{\rm
  shape}^2 + \sigma_{{\rm m},i}^2)$ when the angular separation
between source galaxy $i$ and tracer galaxy $j$ is in angular bin
$\alpha$ and $f'_{ij\alpha} = 0$ otherwise.  Here, $\sigma_{{\rm
    m},i}$ is the shape measurement error of the $i$th source galaxy,
and $\sigma_{\rm shape}=0.22$ is the intrinsic shape noise of the
source galaxies.  Again we use \texttt{TreeCorr} to calculate these
sums.  %\EB{why?}
%Other systematics in the tangential shear measurement
%using DES SV shear catalogs were studied in detail by Clampitt et
%al. (in prep).

$\gamma_T$ is expected to be robust to spatially constant additive
shear systematics since these will cancel when averaging over sources
that are isotropically distributed around tracer galaxies.  This is
one of the main motivations for measuring \gamcor{} rather than
converting $\gamma_T$ to $\kappa$ and performing a $\kappa$-tracer
cross-correlation (the $\gamma$ to $\kappa$ conversion process is not
robust to additive systematics).  Furthermore, by subtracting
$\gamma_T$ measured around random points as in
Eq.~\ref{eq:gammat-est}, we remove the contribution to the tangential
shear measurement from spatially varying additive systematics and edge
effects.

%The estimator of Eq.~\ref{eq:gammat-est} is expected to be robust to
%additive systematic errors in the shear catalog.  Spatially constant,
%additive systematic shears will align nearby sources
%\citep{Mandelbaum:2006a, Clampitt:2015}.  In a Cartesian coordinate
%system oriented along the direction of the systematic this can be
%described by $(\gamma_1, \gamma_2) = (\gamma_{\rm sys}, 0)$, where
%$\gamma_{\rm sys}$ is the systematic-induced shear, and we have
%assumed that the systematic has the same orientation and magnitude at
%every source.  For two sources located at $\phi = 0$ and
%$\phi=90^{\circ}$ (or more generally $\phi'$ and $\phi' +
%90^{\circ}$), their average tangential shear from the systematic is
%$\frac{1}{2}\left(\gamma_{\rm syst} - \gamma_{\rm syst} \right) = 0$.  Since
%real sources are isotropically distributed around the tracer galaxies,
%the systematic does not contribute to \gamcor{}.  Even if there are
%residual additive biases, due to spatial variations in additive
%systematics, subtracting the shear around random points as in
%Eq.~(\ref{eq:gammat-est}) will further reduce the contamination.

We perform both the \kapcor{} and \gamcor{} measurements in
$N_{\theta} = 10$ angular bins logarithmically distributed between
$\theta_{{\rm min}} = 3'$ and $\theta_{{\rm max}} = 50'$.  Our choice
of $\theta_{{\rm max}}$ is set by the size of our jackknife regions
(see below), while the choice of $\theta_{{\rm min}}$ is motivated in
\S\ref{subsubsec:bias_param}.

\begin{figure}
  \includegraphics[width=\columnwidth]{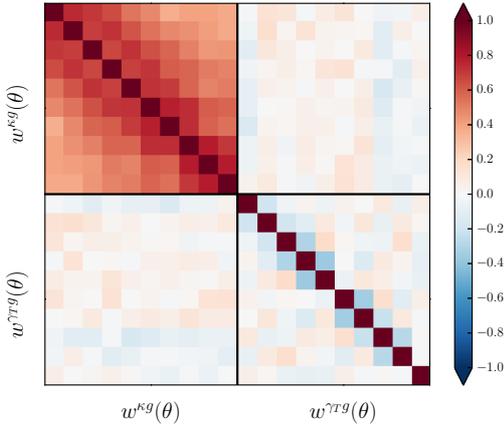}
  \caption{The correlation matrix, $corr(i,j) = C_{ij}/\sqrt{C_{ii}
      C_{jj}}$, for the joint \kapcor{} and \gamcor{} observable,
    where $C_{ij}$ is the covariance matrix element estimate from the
    jackknife.  }
  \label{fig:corr_mat}
\end{figure}

We measure the covariance matrix, $C_{ij}$, of our joint observable
using a jackknife sampling approach; the indices $i$ and $j$ here run
from one to $2N_{\theta}$, the length of our joint data vector.
First, the survey region is divided into $N_{\rm JK} = 200$ roughly
equal-area regions.  The \kapcor{} and \gamcor{} measurements are then
repeated with each of the jackknife regions removed.  The full
covariance matrix for the joint \kapcor{} and \gamcor{} observable is
then computed using the standard jackknife expressions
\citep{Norberg:2009}.  This approach to measuring the covariance has
been validated for \kapcor{} in \citet{Giannantonio:2016}.  Further
tests of the jackknife covariance estimation for \gamcor{} are
presented in \citet{Clampitt:2016}.  The maximum angular scale used in
this analysis ($\theta_{{\rm max}} = 50'$) is chosen to be comparable
to the size of the jackknife subregions.  The correlation matrix,
$corr(i,j) = C_{ij}/\sqrt{C_{ii}C_{jj}}$ computed from the jackknife
covariance, $C_{ij}$, is shown in Fig.~\ref{fig:corr_mat}.  The
observables are ordered as indicated in the figure, with $\theta$
increasing to the right.

Fig.~\ref{fig:corr_mat} shows that \kapcor{} measurements at different
angular scales are significantly correlated.  This is primarily due to
noise correlations in the $\kappa$ map, which arise because the two
main sources of $\kappa$ noise---the Gaussian primary CMB fluctuations
and noise in the SPT temperature maps---are significantly non-white in
the pixel domain.  Additionally, the smoothing of the $\kappa$ maps
with a Gaussian beam (see \S\ref{subsec:kappa_map}) increases the
correlations between different angular scales.  On the other hand, it
is clear from the figure that the \kapcor{} and \gamcor{} measurements
are relatively uncorrelated.  This is due to the fact that the
dominant noise sources in these two observables are uncorrelated.
Noise in \kapcor{} is dominated by noise in the CMB $\kappa$ map which
which receives contributions from both the primary CMB noise and
instrumental noise.  Noise in the measurement of \gamcor{} is
dominated by shape noise at small angular scales, which in turn is
uncorrelated with noise in the CMB $\kappa$ reconstruction.

\section{Analysis}
\label{sec:analysis}

\subsection{Modeling \texorpdfstring{$w^{\kappa g}(\theta)$}{w} and \texorpdfstring{$w^{\gamma_T g}(\theta)$}{w}}

We model \kapcor{} using Eqs.~\ref{eq:wtheta} and \ref{eq:limber},
while \gamcor{} is modeled using Eq.~\ref{eq:limber_gamma}.  As
described in \S\ref{subsec:kappa_map}, the CMB-derived $\kappa$ map is
high-pass filtered to remove modes with $\ell < 30$ and smoothed with
a Gaussian beam of $\theta_{{\rm FWHM}} = 5.4'$.  To make our model
well-matched to the data, we therefore also apply this same filtering
and smoothing to our model for \kapcor{}.  To account for the
high-pass filtering, we set $C_{\ell}^{\kappa g} = 0$ for $\ell < 30$
when computing the sum in Eq.~\ref{eq:wtheta}.  To account for the
beam smoothing, we multiply each term in the sum in
Eq.~\ref{eq:wtheta} by $B_{\ell}^2 = e^{-\ell(\ell+1)\sigma^2}$.

\begin{comment}
The Limber approximation is expected to break down at large angular
scales.  To determine whether our measurements are safely in the
regime for which the Limber approximation holds, we compare our model
prediction to those of a code that does not make the Limber
approximation \citep{Giannantonio:2016}. For angular scales below 50'
(where most of our constraining power comes from), we find agreement
at the sub-percent level.  At the maximum angular scale considered in
this work ($\sim 80'$), agreement between the two codes is better than
3\%, well below the statistical uncertainty in our measurements.  We
are therefore justified in using the Limber approximation throughout. 
\end{comment}

\subsubsection{Bias model}
\label{subsubsec:bias_param}

To fully define our model for the observed lensing cross correlations
in Eqs.~\ref{eq:wtheta}, \ref{eq:limber}, and \ref{eq:limber_gamma},
we must model both the matter power spectrum, $P(k,\chi)$, and the
bias of the tracer galaxies, $b(k, \chi)$.  At scales where the matter
perturbations are in the linear regime, the power spectrum can be
accurately modeled using a Boltzmann code.  To this end, we use
\texttt{CAMB}\footnote{\texttt{http://camb.info}} \citep{CAMBI, CAMBII}.  In
the nonlinear regime, the matter power spectrum is more difficult to
model.  A common prescription is \texttt{Halofit} \citep{Smith:2003},
which calibrates the nonlinear matter power spectrum using dark
matter-only N-body simulations.  Here we use the updated
\texttt{Halofit} model from \citet{Takahashi:2012}.  Although
\texttt{Halofit} has been shown to accurately reproduce the galaxy
power spectrum at moderately nonlinear scales
\citep[e.g.][]{Crocce:2016}, its predictions become more uncertain at
still smaller scales where, for instance, baryonic feedback effects
may become large \citep[e.g.][]{Jing:2006, vanDaalen:2011}.  At some
level, uncertainty in the nonlinear matter power spectrum is
incorporated into our analysis through our modeling of the tracer
bias, as we discuss in more detail below.

A common approach to parameterizing $b(k,\chi)$ is the so-called
linear bias model \citep{Mo:1996, Matarrese:1997}, for which the bias
has no scale dependence, but is allowed to vary with comoving
distance: $b(k,\chi) = f(\chi)$.  It is well known that the linear
bias model accurately describes galaxy clustering over scales where
the matter density perturbations are linear, and even at scales
several times smaller than the transition scale from the linear to the
nonlinear regime \citep[e.g.][]{Crocce:2016}.  However, at still
smaller scales, the simple linear bias model is expected to break
down. Since small angular scales contain useful information about the
lensing cross-correlations, we would like to use the smallest scale
possible for which we can still develop a reasonable bias model.
However, we do not know the exact scale at which linear bias remains
valid.  We therefore take the approach of choosing a range of angular
scales for which we believe linear bias to be a reasonable
approximation, but introduce additional freedom into our bias model to
capture small departures from linear bias.  Furthermore, by allowing
for freedom in the bias model at small scales, we effectively account
for uncertainty in the matter power spectrum at these scales due, for
instance, to baryonic effects.

For the benchmark sample used in this work, \citet{Crocce:2016} have
determined that the linear bias approximation begins to break
down at angular scales $\lesssim 3'$ in their measurements of the
galaxy auto-correlation at $0.4 < z < 0.6$.  We therefore adopt
$\theta_{{\rm min}} = 3'$ in this work.  Since \citet{Crocce:2016}
find that the linear bias approximation is valid at scales smaller
than $3'$ for $z > 0.6$, this is a somewhat conservative choice.

We allow for freedom beyond linear bias with a second order Taylor
expansion in $k$ and $z$:
\begin{multline}
\label{eq:bias_model}
b(k,z) = b_0\left[1 + a_1 (k/k_0) + a_2 (k/k_0)^2 \right] \\
 \times \left[1 + c_1 (z/z_0) + c_2 (z/z_0)^2  \right],
\end{multline}
where $b_0$, $a_i$ and $c_i$ are parameters of the model.  Many
different approaches to parameterizing the scale dependence
\citep[e.g.][]{Bielefeld:2015} and redshift dependence
\citep[e.g.][]{Fry:1996, Matarrese:1997, Clerkin:2015} of the bias
have been proposed in the literature.  Since we are only attempting to
capture small deviations from linear bias, a Taylor expansion in $k$
is appropriate here.  Our assumed form for the redshift dependence of
the bias is simple and flexible.  As we show in
\S\ref{subsec:bias_marg}, our constraints on the redshift dependence
of the bias are weak, so the precise form adopted for this dependence
is relatively unimportant to our analysis.

We choose $k_0 = 1\, h\,{\rm Mpc}^{-1}$ for the pivot scale since this
is roughly where we expect linear bias to begin breaking down
\citep{Crocce:2016}, and $z_0 = 0.6$ since this is near the center of
the redshift distribution of the tracer galaxies.  At $z=0.6$, $1/k =
1\,h^{-1}{\rm Mpc}$ corresponds to an angular scale of $3.6'$,
slightly greater than the minimum angular scale, $\theta_{{\rm min}} =
3'$.  We place flat priors with range $[-5, 5]$ on the bias
parameters $a_i$ and $c_i$; in our analysis we find these priors are
sufficiently wide that they have no effect on any of our constraints.

\subsection{Model Parameterizations}

Given a cosmological model and the bias model of
Eq.~\ref{eq:bias_model}, we can compute \kapcor{} and \gamcor{} using
the measured $N(z)$ of the tracer and source galaxies.  As discussed
above, we expect the joint measurement of these two quantities to be
sensitive to cosmological parameters, systematics in the lensing
measurements, and systematics in the galaxy redshift distributions.
We now introduce three model parameterizations that are chosen to
explore our sensitivity to these quantities.

\subsubsection{Cosmology Analysis}
\label{subsubsec:cosmo_param}

We are interested in the sensitivity of our joint measurement of
\kapcor{} and \gamcor{} to cosmological parameters.  As discussed in
\S\ref{sec:formalism}, we expect the joint measurement to be
particularly well suited to constraining cosmological parameters that
change the geometry or expansion history of the Universe.  In the
interest of simplicity, we focus on flat $\Lambda$CDM cosmological
models and explore our ability to constrain $\Omega_M$.  All other
cosmological parameters are fixed to their best-fit values from a
flat-$\Lambda$CDM-model fit by \citet{PlanckXIII:2015} (hereafter
\citetalias{PlanckXIII:2015}) to their observations of the CMB.  These
parameter values are $h = 0.6751$, $\Omega_b = 0.0488$, $\tau =
0.063$, $n_S = 0.9653$, and $A_S = 2.1306\times 10^{-9}$ at a pivot
scale of $k = 0.05\,{\rm Mpc}^{-1}$, corresponding to $\sigma_8 =
0.815$.

The cosmological analysis considered here is optimistic in the sense
that we only vary $\Omega_M$ and marginalize over the bias parameters
$a_i$ and $c_i$.  A complete cosmological analysis should also
marginalize over additional cosmological parameters.  Given that our
constraints on $\Omega_M$ are relatively weak compared to those from
\citetalias{PlanckXIII:2015}, including the uncertainties on the
best-fit cosmological parameters from \citetalias{PlanckXIII:2015}
would have only a small impact on our results.  We remind the reader
that the analysis presented here is concerned with early DES SV data
and should be viewed as a proof-of-principle work that demonstrates
the potential of similar analyses with future DES and SPT data.  The
cosmological parameterization and the associated priors are summarized
in Table~\ref{tab:params}.

\subsubsection{Shear bias parameterization}
\label{subsubsec:shear_bias_param}

The joint measurement of \kapcor{} and \gamcor{} is sensitive to
systematics in the measurements of the CMB-derived $\kappa$ and the
galaxy-lensing derived $\gamma_T$.  We focus on systematics in the
$\gamma_T$ estimates.  Systematics affecting the measurement of galaxy
shear can result from a number of sources, including incomplete
modeling of telescope optics, atmospheric distortion and contamination
from nearby sources \citepalias{Jarvis:2015}.

%Systematics in shear
%estimation are commonly parameterized with an additive term, $c$, and
%a multiplicative term, $m$, such that the measured shear,
%$\gamma^{{\rm obs}}$, is related to the true shear, $\gamma^{{\rm
%    true}}$, by $\gamma^{{\rm obs}} = (1+m)\gamma^{{\rm true}} + c$
%\citep{Heymans:2006, Huterer:2006}.

We investigate the ability of the joint \kapcor{} and \gamcor{}
measurement to constrain systematics that scale the absolute
calibration of the measured shear, i.e. {\it multiplicative biases}.
We have two reasons for focusing on multiplicative biases and ignoring
systematics that additively bias the measured shear.  First, as
discussed in \S\ref{sec:measurement}, our measurement of \gamcor{} is
expected to be robust to additive shear systematics.  Second, joint
measurement of \gamcor{} and \kapcor{} is particularly well suited to
constraining multiplicative shear systematics \citep{Vallinotto:2013}.
If one only measures \gamcor{}, the multiplicative systematic will be
completely degenerate with the scale-independent tracer bias since
both affect the normalization of \gamcor{}.  Joint measurement with
\kapcor{}, however, allows this degeneracy to be broken since
\kapcor{} does not depend on galaxy shear systematics at all.

Typically, systematic errors in galaxy shear measurements are
estimated by dividing the full shear catalog into subsamples that are
expected to yield consistent shear estimates, but may not because of
some systematic effect (e.g. \citealt{Mandelbaum:2006b}).  The
difference between shear estimates from the various subsamples is then
reflective of the magnitude of the systematic effect in question.
However, these internal consistency tests do not constrain the
absolute shear calibration since systematic errors in the calibration
will affect all subsamples in the same way.  Therefore, to estimate
errors on the shear calibration, one often relies on image simulations
\citep[e.g.][]{Schrabback:2007,Kacprzak:2012}.  Image simulations,
however, can be problematic if they are not perfectly matched to the
datasets in question.

The joint measurement of \kapcor{} and \gamcor{} considered here can
be used to constrain the absolute shear calibration.  Multiplicative
systematics in the galaxy shears will not affect the CMB lensing
measurements, allowing the degeneracy between shear calibration and
galaxy bias to be broken in the joint lensing measurement.  Therefore,
the joint lensing measurement can be used to constrain the absolute
shear calibration without resorting to simulations.

Following the convention of \citet{Heymans:2006}, we parameterize the
multiplicative galaxy shear systematic with a single parameter, $m$,
so that the model tangential shear is
\begin{eqnarray}
\gamma_T = (1+m) \gamma_T^{{\rm no-sys}},
\end{eqnarray}
where $\gamma^{{\rm no-sys}}$ is the model shear in the
absence of any systematic effect.  In the absence of any systematic
effect we should recover $m = 0$.

\subsubsection{Redshift bias parameterization}
\label{subsubsec:redshift_bias_param}

We also expect the joint lensing measurement considered here to be
sensitive to systematic errors in the photometric redshift estimates
of the tracer and source galaxies.  Systematics in the photo-$z$
estimates of the tracer galaxies will change the model predictions for
both \kapcor{} and \gamcor{}; systematics in the photo-$z$ estimates
of the source galaxies will change the predicted \gamcor{}, but not
the predicted \kapcor{}.  Therefore, it should be possible to
constrain photo-$z$ systematics in the joint fit to these two
lensing-tracer correlations.  Taking this reasoning a step further, in
principle one could use the joint \kapcor{} and \gamcor{} measurement
to constrain the full $N(z)$ for the tracer or source galaxies.  For
this first measurement with DES SV data, however, we expect the
signal-to-noise to be relatively low.  We therefore focus here on
constraining a single photo-$z$ systematic parameter rather than the
full $N(z)$.  Future DES data will make constraining the full $N(z)$
with joint measurement of \kapcor{} and \gamcor{} an exciting
possibility (see discussion in \S\ref{sec:discussion}).
 
It is easiest to gain intuition for the ability of the joint lensing
observable to constrain the tracer and source galaxy redshift
distributions in the limit that the tracer redshift distribution is
narrow and the bias is scale and redshift-independent.  In that limit,
$\kapcormath{} \propto g^{{\rm CMB}}(\chi^g) b_0$ and $\gamcormath{}
\propto g^{s}(\chi^g) b_0$, where the $g$-factors are defined in
\S\ref{sec:formalism}, $\chi^g$ is the comoving distance of the tracer
galaxies, and $b_0$ is the tracer bias.  The measured redshift
distributions then enter into the model predictions for \kapcor{} and
\gamcor{} only through $g^{{\rm CMB}}(\chi^g)$ and $g^{s}(\chi^g)$.
For a single measurement of either \kapcor{} or \gamcor{}, then, $b_0$
is completely degenerate with the redshift information.  The joint
lensing cross-correlation measurement, however, breaks this degeneracy
because $g^{{\rm CMB}}(\chi^g)$ and $g^{s}(\chi^g)$ depend in
different ways on the tracer and source redshift distributions.

Photometric redshift estimation is a notoriously difficult problem,
and can be affected by a host of different systematic errors (in the
context of DES see, for instance, \citealt{Sanchez:2014}).  These
systematic errors can change the inferred $N(z)$ of both the sources
and the tracers in complicated ways.  Here, we take a simplistic
approach and parameterize only the systematic error in the tracer
photometric redshifts with a single parameter, $\Delta_z$, which
simply shifts (in redshift) the model $N(z)$ for the tracers:
\begin{eqnarray}
N(z) =
\begin{cases}
 N_{{\rm no-sys}}(z - \Delta_z), & \text{if } z -\Delta_z > 0 \\
 0, & \text{otherwise},
\end{cases}
\end{eqnarray}
where $N_{{\rm no-sys}}(z)$ is the tracer $N(z)$ in the absence of any
systematic.  Note that we enforce the physical requirement that $N(z)
= 0$ for $z < 0$.  The treatment adopted here has the advantages of
simplicity and generality: any systematic which changes the mean
$N(z)$ of the tracers is likely to generate an effective $\Delta_z$,
and should therefore be constrained by this analysis.
\citet{DESshear2pt} have adopted the same parameterization of
photo-$z$ systematics in the analysis of the cosmic shear two-point
function.  Note that although we consider only $\Delta_z$ for the
tracers in this analysis, the joint analysis of \gamcor{} and
\kapcor{} could also be used to constrain $\Delta_z$ for the sources.
We have chosen to focus on $\Delta_z$ for the tracers because the
constraint on $\Delta_z$ for the sources is quite weak with current
data.

\subsubsection{Other sources of systematic error}

Of course, there are many other ways that systematics could affect the
joint lensing-galaxy cross-correlation measurements besides those
considered above.  We focus our analysis on multiplicative shear
systematics and photometric redshift systematics because these are
likely to be some of the most significant sources of systematic error
in the data, and because the joint \kapcor{} and \gamcor{} measurement
is particularly well suited to constraining these systematics.

\citet{Crocce:2016} have constrained several different potential
sources of systematic error --- including variations in observation
conditions and stellar and dust contamination --- that may impact the
distribution of benchmark galaxies and have found that above $z
\gtrsim 0.4$, their impact is likely small.  Several systematic
effects that may affect the SPT $\kappa$ maps have been considered by
\citet{vanEngelen:2012}.  These include sources of contamination, such
as emissive point sources, the Sunyaev-Zel'dovich effect, and the
cosmic infrared background, as well as other effects, such as beam
uncertainties.  The analysis of \citet{vanEngelen:2012} indicates that
the impact of these sources of systematic error on their measurement
of the lensing power spectrum is significantly smaller than the
corresponding statistical errorbars, suggesting that such effects
likely have a negligible impact on the analysis presented here.
\citet{Giannantonio:2016} have considered how several different
systematics --- in both DES and SPT data --- may impact the
measurement of \kapcor{}, finding no evidence for significant
contamination. Finally, we mention that if there is overlap between the
source and tracer galaxies in redshift, then intrinsic alignment
effects \citep{Troxel:2015} can lead to a distinct signature in the
measured $\gamma_T$ that could in principle be constrained using the
joint \kapcor{} and \gamcor{} observable.  Since \gamcor{} would be
affected by intrinsic alignments while \kapcor{} is not, the joint
lensing observable is a potentially attractive probe of these effects.

\subsection{Likelihood}

We adopt a Gaussian likelihood for the data given our model vector:
\begin{eqnarray}
\label{eq:likelihood}
\mathcal{L}(\vec{d} | \vec{p}) \propto \exp \left[-\frac{1}{2}\left(\vec{d} - \vec{t}(\vec{p})\right)^T \widehat{\mathbf{C}^{-1}} \left(\vec{d}-\vec{t}(\vec{p})\right) \right],
\end{eqnarray}
where $\vec{d} = (\hat{w}^{\kappa g}(\vec{\theta}), \hat{w}^{\gamma_T
  g}(\vec{\theta}))$ is the joint data vector of the CMB and galaxy
lensing measurements and $\vec{t}(\vec{p})$ is the model (theory)
vector, expressed as a function of the parameter vector, $\vec{p}$.
$\widehat{\mathbf{C}^{-1}}$ is our estimator for the inverse
covariance matrix of the data vector.  Following \citet{Hartlap:2007}
and \citet{Friedrich:2016}, we estimate the inverse covariance matrix
using
\begin{eqnarray}
\label{eq:inv_cov_estimation}
\widehat{\mathbf{C}^{-1}} = \frac{N - d - 2 }{N-1} \mathbf{C}^{-1},
\end{eqnarray}
where $N$ is the number of jackknife regions (in this case $N= 200$),
$d$ is the length of our data vector (in this case $d = 20$) and
$\mathbf{C}$ is the covariance matrix estimated from the jackknifing
procedure.  

We parameterize $\vec{t}(\vec{p})$ as discussed in
\S\ref{subsubsec:cosmo_param}, \S\ref{subsubsec:shear_bias_param}, and
\S\ref{subsubsec:redshift_bias_param}.  For each of these three
parameterizations, we hold the parameters in the other two models
fixed.  For the cosmology analysis this means fixing $m = 0$, and
$\Delta_z = 0$.  For the shear and redshift systematics analyses, this
means fixing the cosmological model to the best fit flat-$\Lambda$CDM
cosmological model from the CMB only analysis of
\citetalias{PlanckXIII:2015} and fixing $\Delta_z=0$ or $m=0$,
respectively.  These values (and the other priors imposed on our
models) are summarized in Table~\ref{tab:params}.  This approach is
motivated by the two main goals of this work. First, we wish to show
that the measurements of \kapcor{} and \gamcor{} are self consistent
and that they agree with the currently favored flat-$\Lambda$CDM
cosmological model.  For this purpose, it is sufficient to consider
the parameter constraints along particular directions in parameter
space.  Second, we wish to highlight the potential of the joint
\kapcor{} and \gamcor{} measurements to constrain cosmology, shear
systematics and redshift distributions.  With current DES SV data, the
constraints that we obtain on the model parameters are weak relative
to other published constraints.  Treating each model fit separately,
then, can be seen as imposing tight (but realistic) external priors on
the quantities that are not varied.  Finally, the approach adopted
here has the advantage of simplicity, appropriate for a first
measurement that has low signal-to-noise.  For future joint
measurements of \kapcor{} and \gamcor{} that have higher
signal-to-noise it may be appropriate to vary all of the model
parameters simultaneously.

Given the likelihood of Eq.~\ref{eq:likelihood} and the priors
discussed above, we can calculate the posterior on our model
parameters.  We sample the multidimensional posterior using a Markov
Chain Monte Carlo approach implemented with the code \texttt{emcee}
\citep{Foreman-Mackey:2013}.  Our entire pipeline (from computation of
the model vector to sampling of the parameter space) is implemented
using
\texttt{COSMOSIS}\footnote{\texttt{https://bitbucket.org/joezuntz/cosmosis/wiki/Home}}
\citep{Zuntz:2015}.

\section{Results}
\label{sec:results}

\begin{figure*}
  \includegraphics[width=2\columnwidth]{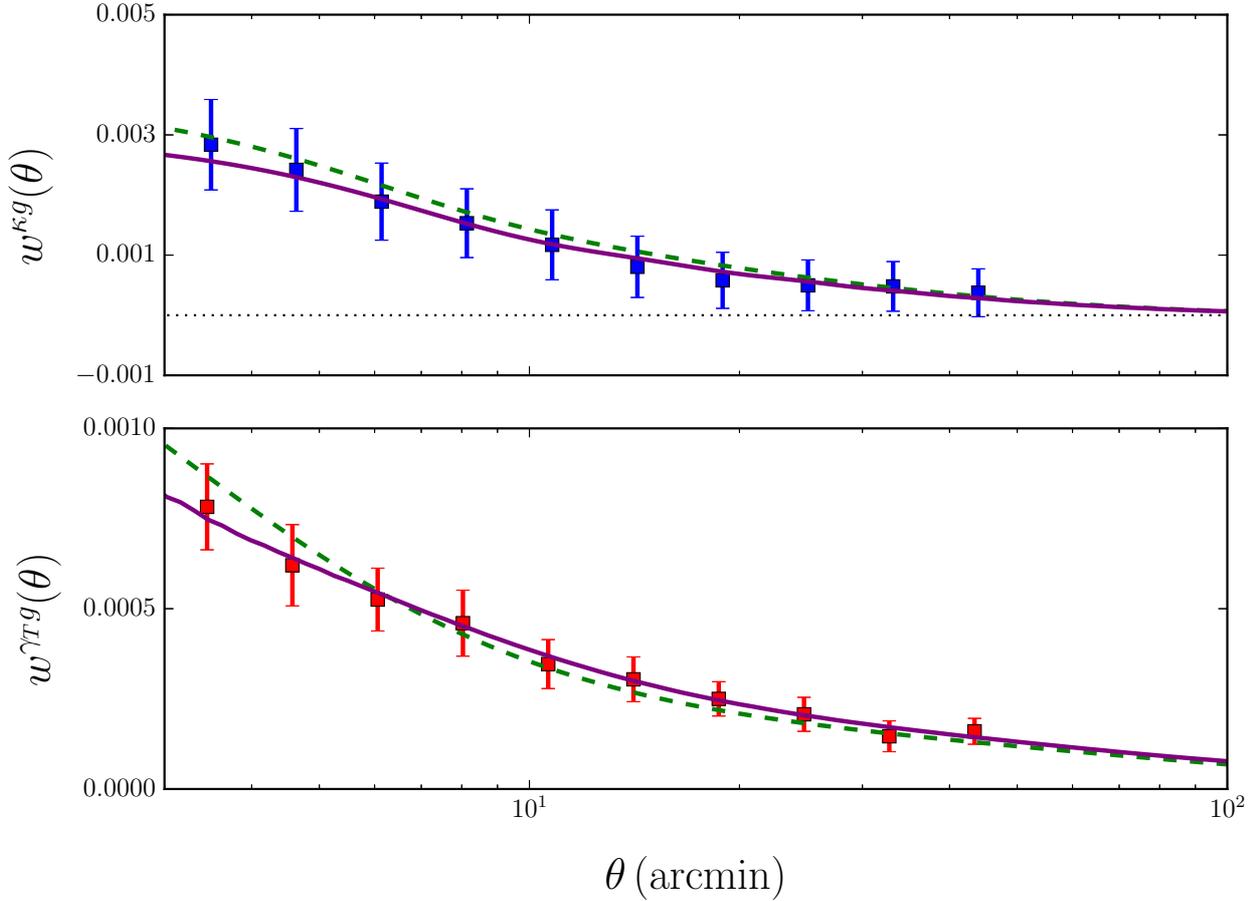}
  \caption{The measured \kapcor{} (top panel) and \gamcor{} (bottom
    panel).  Error bars are the square roots of the diagonal elements
    of the covariance matrix.  Solid (purple) curve shows the best-fit
    model using the parameterization discussed in
    \S\ref{subsubsec:cosmo_param}, for which $\Omega_M$ and $b_0$,
    $a_i$ and $c_i$ are free parameters.  Dashed (green) curve shows
    the best-fit model when only $b_0$ is allowed to vary,
    $a_i=c_i=0$, and the cosmological parameters are fixed to the
    best-fit flat-$\Lambda$CDM model from the CMB-only analysis of
    \citetalias{PlanckXIII:2015}.  Both model curves result from {\it
      joint} fits to the \kapcor{} and \gamcor{} measurements.  The
    dotted line in the top panel shows $\kapcormath{} = 0$.}
  \label{fig:data}
\end{figure*}

%\begin{figure}
%  \includegraphics[width=\columnwidth]{bias_compare.eps}
%  \caption{Posteriors on the linear bias parameter, $b_0$, for the
%    tracer galaxies resulting from fits to the \kapcor{} (blue solid)
%    and \gamcor{} (green dashed) measurements. In these fits, the
%    cosmological model is fixed to the best-fit flat-$\Lambda$CDM
%    model from the CMB-only analysis of \citetalias{PlanckXIII:2015}.
%    The two measurements of the tracer $b_0$ are consistent with each
%    other.}
%  \label{fig:bias_compare}
%\end{figure}

\subsection{Measurement and consistency test}
\label{subsec:b0fits}

Fig.~\ref{fig:data} shows our measurements of \kapcor{} (top panel)
and \gamcor{} (bottom panel). The error bars shown are the diagonal
elements of the jackknife covariance matrix.  Note that the error bars
on both the \kapcor{} and \gamcor{} measurements are correlated
between different angular scales, as shown in Fig.~\ref{fig:corr_mat}.
The significance of the joint \kapcor{} and \gamcor{} measurement is
roughly $19\sigma$.  The solid (purple) curve in Fig.~\ref{fig:data}
represents the best-fit model from the cosmology analysis discussed in
\S\ref{subsubsec:cosmo_param}, where $\Omega_M$ and the bias
parameters $b_0$, $a_i$, and $c_i$ are allowed to vary.  Note that
this curve represents a {\it joint} fit to the \kapcor{} and \gamcor{}
measurements; in other words, the same parameters define the model
curves in both the top and bottom panels.  The dashed (green) curve in
Fig.~\ref{fig:data} represents the best-fit model when we fix
$\Omega_M = 0.3121$ (i.e. the best fit value from a flat-$\Lambda$CDM
fit to CMB observations in \citetalias{PlanckXIII:2015}) and $a_i =
c_i = 0$, but allow $b_0$ to vary.  Both model curves agree well with
the data, and neither model is strongly preferred over the other (as
we quantify in more detail below).  The dotted (black) curve in
Fig.~\ref{fig:data} corresponds to $\kapcormath{} = 0$.

As a consistency check on the data, we first perform fits to the
\kapcor{} and \gamcor{} measurements separately (i.e. {\it not}
jointly) in which the cosmological model is fixed to the best fit
flat-$\Lambda$CDM model from the analysis of CMB data by
\citetalias{PlanckXIII:2015}, while the redshift-independent bias
coefficient, $b_0$, is allowed to vary and $a_i = c_i = 0$.  These
fits are not the primary focus of this work (because they treat the
\kapcor{} and \gamcor{} measurements separately and because they
assume constant bias down to the smallest angular scales that we
measure) but they allow us check the SPT and DES data for consistency,
and also to compare how constraining these two data sets are.  From
the analysis of \kapcor{}, we find $b_0 = 1.14\pm0.31$; from the
analysis of \gamcor{} we find $b_0 = 1.26\pm0.07$.  Making the
assumption that the errors on the \kapcor{} and \gamcor{} measurements
are uncorrelated (a reasonable assumption given
Fig.~\ref{fig:corr_mat}), these two $b_0$ constraints are consistent
at roughly $0.4\sigma$.  However, the constraints on $b_0$ from
\gamcor{} are tighter than those from \kapcor{} by roughly a factor of
4.6. This is because the error bars on \kapcor{} are larger and
  more correlated than the error bars on \gamcor{}. The tightness of
the joint constraint on $b_0$ suggests that we may be able to measure
variation of the bias with redshift; this possibility is explored more
in \S\ref{subsec:bias_marg}.

The constraints on $b_0$ obtained above are also consistent with the
results of other analyses of DES SV data.  Our measurement of $b_0$
from SPT data is consistent with the measurements of
\citet{Giannantonio:2016}, which also measured \kapcor{} using a
similar catalog of benchmark galaxies and identical CMB $\kappa$ maps.
\citet{Giannantonio:2016} found best fit constant biases of $b_0 =
0.75\pm0.25$ and $b_0 = 1.25 \pm 0.25$ for the redshift bins $z \in
[0.4, 0.6]$ and $z \in [0.6, 0.8]$, respectively.  Averaging these two
measurements yields a bias of $b_0 \sim 1.0$, consistent with our
measurement of $b_0$ using \kapcor{} within the error bars.  A
quantitative comparison to \citet{Giannantonio:2016} is difficult
because we use an identical CMB-$\kappa$ map, but somewhat different
versions of the benchmark catalog (and mask) and a different
photometric redshift estimation code.

\begin{figure*}
  \includegraphics[width=2\columnwidth]{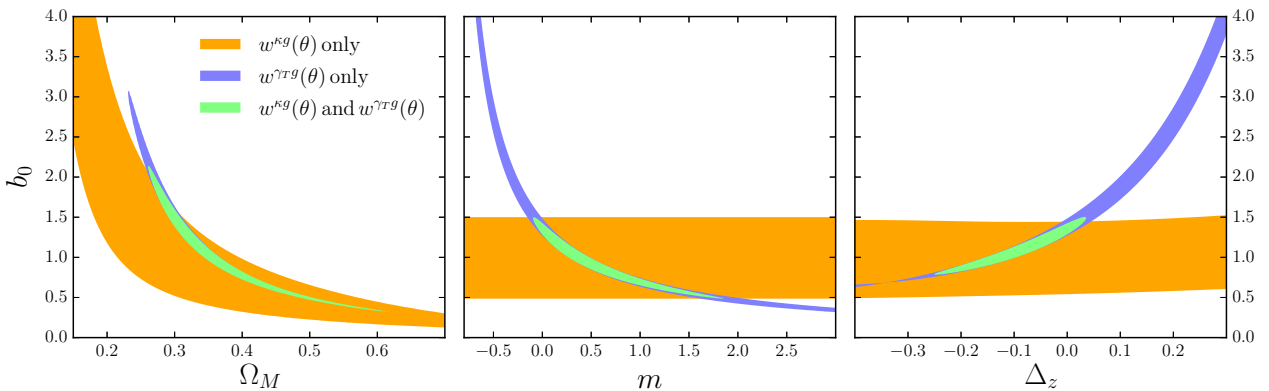}
  \caption{Constraints obtained on the three model parameters when we
    fix the bias parameters $a_i = c_i = 0$, but allow $b_0$ to vary;
    contours show where $\Delta \chi^2 = 1$ relative to the minimum
    $\chi^2$.  Orange contour shows the constraint obtained from
    analysis of \kapcor{} alone; blue contour shows the constraint
    obtained from analysis of \gamcor{} alone; green contour shows the
    constraint obtained from the joint analysis of \kapcor{} and
    \gamcor{}.  In all cases, the joint measurement of \kapcor{} and
    \gamcor{} helps to break degeneracies between the model parameters
    of interest and the bias parameter $b_0$.  We have restricted the
    analysis here to angular scales $\theta > 10'$ to ensure that
    linear bias remains valid.}
  \label{fig:constraints_2d}
\end{figure*}

Our measurement of $b_0$ is also consistent with constraints on $b_0$
obtained by \citet{Crocce:2016} from measuring galaxy clustering of
the benchmark galaxy sample.  Using the TPZ photometric redshift
algorithm \citep{CarrascoKind:2013,CarrascoKind:2014},
\citet{Crocce:2016} found best fit constant bias parameters $b_0 =
1.29\pm0.04$ and $b_0 = 1.34\pm0.05$ for the redshift bins $z\in [0.4,
  0.6]$ and $z \in [0.6, 0.8]$, respectively.  These two measurements
yield an average bias of $b_0 \sim 1.32$, which is consistent with the
constraints we obtain from \kapcor{} and \gamcor{} to within the error
bars.  Again, a quantitative comparison to \citet{Crocce:2016} is
difficult because we use a slightly different benchmark catalog and a
different photometric redshift estimation code.
%Our measurement of $b_0$ is also consistent at the $\sim1\sigma$ level
%with the results of Chang et al., in prep, which will present a
%constraint on the bias of the benchmark galaxies using a technique
%developed in \citet{Amara:2012}.

\subsection{Bias degeneracies}
\label{subsec:bias_degeneracies}

To gain intuition for how the joint \kapcor{} and \gamcor{}
measurement breaks degeneracies with tracer bias, we now present
constraints in the two dimensional plane defined by $b_0$ and each of
the three model parameters defined in \S\ref{subsubsec:cosmo_param},
\S\ref{subsubsec:shear_bias_param}, and
\S\ref{subsubsec:redshift_bias_param}.  For this analysis, we fix the
bias parameters $a_i = c_i = 0$, which corresponds to a constant bias
model described by $b_0$ alone; doing so considerably simplifies the
interpretation and visualization of the results.  However, as noted
previously, we expect that constant bias may not accurately describe
the data at small angular separations.  We therefore restrict the
analysis presented in this section to angular scales $\theta > 10'$,
which should be safely in the linear bias regime \citep{Crocce:2016}.
Imposing this restriction on the data will weaken our constraints, but
we remind the reader that our intent in this section is only to gain
intuition for degeneracies with $b_0$.  In
\S\ref{sec:bias_marg_results} we will present results that use data at
small angular scales and for which we allow the bias parameters $a_i$
and $c_i$ to vary.

The leftmost panel of Fig.~\ref{fig:constraints_2d} presents the
constraints obtained from the analysis of \kapcor{} and \gamcor{} in
the $\Omega_M$--$b_0$ plane.  Each shaded region corresponds to a
contour of the posterior probability such that the $\Delta\chi^2$
relative to the minimum is $\Delta \chi^2 = 1$ (this value of $\Delta
\chi^2$ was chosen to improve the visualization since the constraints
obtained in this analysis are fairly weak owing to the exclusion of
the small angle measurements).  The orange region shows the
constraints obtained from analysis of \kapcor{} alone; the blue region
shows the constraints obtained from analysis of \gamcor{} alone; the
green region shows the constraints obtained from the joint analysis of
\kapcor{} and \gamcor{}.  Since there is little covariance between
\kapcor{} and \gamcor{}, the joint constraints are roughly the product
of the individual constraints.  From the figure it is clear that there
is a strong degeneracy between $\Omega_M$ and $b_0$ for both \kapcor{}
and \gamcor{}.  The joint measurement of both \kapcor{} and \gamcor{}
helps to break this degeneracy.

The middle panel of Fig.~\ref{fig:constraints_2d} shows the
constraints obtained from the analysis of \kapcor{} and \gamcor{} in
the $m$--$b_0$ plane.  Since \kapcor{} does not depend at all on $m$, we
obtain no constraint on $m$ from the analysis of \kapcor{} alone
(orange region).  \gamcor{} depends on $m$, but in a way that is
completely degenerate with $b_0$ (blue region); we therefore also
obtain no constraint on $m$ from \gamcor{} alone.  The joint
measurement of \kapcor{} and \gamcor{}, however, breaks this
degeneracy with the bias as shown by green region.

The rightmost panel of Fig.~\ref{fig:constraints_2d} shows the
constraints obtained from the analysis of \kapcor{} and \gamcor{} in
the $\Delta_z$--$b_0$ plane.  Changing $\Delta_z$ does not have a very
large impact on \kapcor{} because the CMB source plane is at much
higher redshift than the tracer galaxies; any reasonable $\Delta_z$ is
very small compared to the redshift of the surface of last scattering.
This fact combined with the low signal-to-noise of the \kapcor{}
measurement means that we do not obtain a constraint on $\Delta_z$
from \kapcor{} alone (orange region).  Furthermore, because the
constraint obtained from \gamcor{} alone is highly degenerate with
$b_0$, we also do not obtain a constraint on $\Delta_z$ from \gamcor{}
alone (blue region).  The joint measurement of \kapcor{} and
\gamcor{}, however, breaks the degeneracy between $\Delta_z$ and $b_0$
(green region). 

It is also interesting to note the direction of the degeneracy between
between $\Delta_z$ and $b_0$.  For the constraints obtained from
\gamcor{}, there is a clear positive correlation between $\Delta_z$
and $b_0$.  This is because increasing $\Delta_z$ for the tracer
galaxies pushes $W^g(z)$ to higher redshift.  At high redshift,
$g^s(z)$ is reduced and $P(k,z)$ is also reduced because of the growth
of structure.  These two effects lead to a lower \gamcor{}, which is
offset by increasing $b_0$.  The net result is a positive correlation
between $\Delta_z$ and $b_0$.  Somewhat suprisingly, there is also a
very slight positive correlation between $\Delta_z$ and $b_0$ for the
constraints obtained from \kapcor{}.  This slight correlation is due
to two competing effects: $g^{\rm CMB}(z)$ increases with redshift
while $P(k,z)$ decreases with redshift.  For our particular tracer
galaxy sample, the effect of $P(k,z)$ decreasing with redshift
dominates over the effect of $g^{\rm CMB}(z)$ decreasing with
redshift.  The net result is that \kapcor{} decreases slightly with
$\Delta_z$, leading to a weak correlation between $\Delta_z$ and
$b_0$.

\begin{figure}
  \includegraphics[width=1.02\columnwidth]{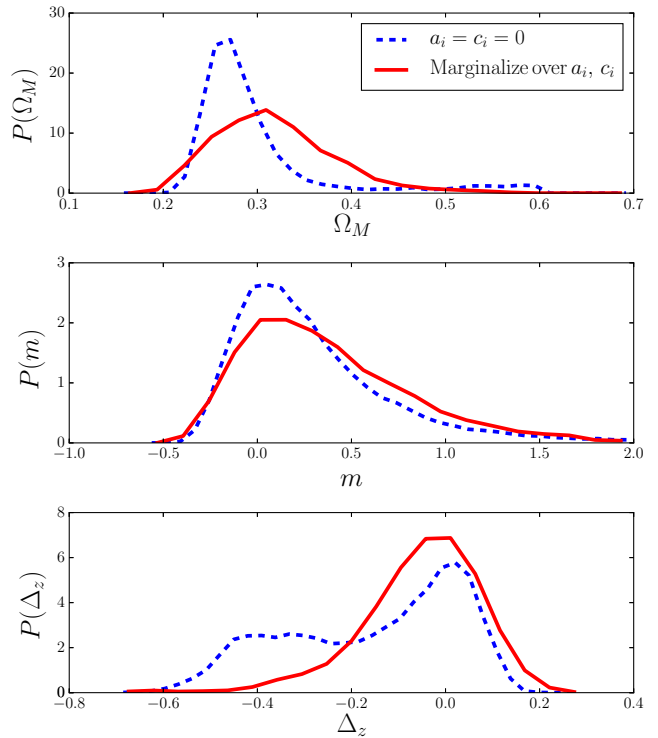}
  \caption{Posteriors on the main parameters of the three models
    discussed in \S\ref{sec:analysis}.  Solid (red) curves show the
    results when we marginalize over the bias parameters $a_i$ and
    $c_i$.  Dashed (blue) curves show the result when we fix $a_i =
    c_i = 0$.  The three analyses are consistent with the best-fit
    $\Lambda$CDM model from \citetalias{PlanckXIII:2015} and with no
    systematic contamination of the shear or photo-$z$ measurements.}
  \label{fig:main_constraints}
\end{figure}

\subsection{Bias-marginalized parameter constraints}
\label{sec:bias_marg_results}

The results presented in \S\ref{subsec:bias_degeneracies} were
restricted to constant bias (i.e. $a_i = c_i = 0$) and for this reason
used only the correlation function measurements at large angular
scales ($\theta > 10'$).  As we have argued in
\S\ref{subsubsec:bias_param}, by allowing additional freedom in our
bias model, we can use measurements at smaller angular scales and
thereby increase our signal-to-noise without worrying about biasing
our results.  We now present the constraints obtained when we allow
$a_i$ and $c_i$ to vary in our model fits (we refer to this as the
{\it evolving bias} analysis).  For these results, we marginalize over
all the bias parameters ($b_0$, $a_i$ and $c_i$), showing only the
posteriors on the model parameters of interest.  The posteriors for
the three analyses of \S\ref{subsubsec:cosmo_param},
\S\ref{subsubsec:shear_bias_param}, and
\S\ref{subsubsec:redshift_bias_param} are shown as the solid (red)
curves in Fig.~\ref{fig:main_constraints}.  For comparison, we also
show the posteriors on model parameters for the constant bias analysis
with $a_i = c_i = 0$ (dashed blue curves).  Both the evolving bias and
the constant bias curves shown in Fig.~\ref{fig:main_constraints} were
obtained using the \kapcor{} and \gamcor{} measurements across the
full angular range shown in Fig.~\ref{fig:data}.

The top panel of Fig.~\ref{fig:main_constraints} shows the posteriors
on $\Omega_M$ resulting from the analysis described in
\S\ref{subsubsec:cosmo_param}.  The 68\% posterior interval on
$\Omega_M$ from the evolving bias analysis is $(0.26,0.39)$, while the
posterior from the constant bias analysis is $(0.25, 0.33)$.  Both of
these intervals contain the best-fit value of $\Omega_M = 0.3121$ from
the CMB-only analysis of \citetalias{PlanckXIII:2015}.  The
constraints from the constant and evolving bias analyses are also
consistent with each other, although the constraint from the evolving
bias analysis is weaker.  The consistency of the results from the two
different bias parameterizations suggests that there is no strong
evidence for departures from linear bias; we quantify this statement
in more detail in \S\ref{subsec:bias_marg}.

Note that our roughly 20\% constraint on $\Omega_M$ uses only
$\sim3\%$ of full survey DES Data.  With more data, we expect the
constraints on \kapcor{} and \gamcor{} to tighten significantly.  This
measurement with early DES SV and SPT data indicates that the data are
in good agreement with each other and with the best fit cosmological
model from \citetalias{PlanckXIII:2015}.  We refrain from comparing
our constraint on $\Omega_M$ to other published constraints because we
have (intentionally) not performed a complete cosmological analysis
here.

The middle panel of Fig.~\ref{fig:main_constraints} shows the
posterior on $m$ --- the multiplicative shear systematic parameter ---
in the evolving bias parameterization.  The dashed (blue) curve shows
the posterior on $m$ for the constant bias parameterization.  The 68\%
posterior interval on $m$ is $(-0.05, 0.80)$ in the evolving bias
parameterization and $(-0.07, 0.68)$ in the constant bias
parameterization.  Both of these intervals contain $m=0$ and are
therefore consistent with no systematic bias of the galaxy shear
measurements.  However, both posteriors on $m$ are highly
non-Gaussian, exhibiting longs tail to quite large values of $m$.

It is interesting to note that unlike the constraints on $\Omega_M$,
the constraints on $m$ appear to be fairly robust to marginalization
over the evolving bias parameters.  This behavior can be understood as
follows.  Information about $\Omega_M$ comes from both the relative
amplitudes of \kapcor{} and \gamcor{} (i.e. the lensing ratio of
\citealt{Das:2009}) as well as the shape of the two correlation
functions.  Marginalization over the evolving bias parameters
effectively washes out some of the information content in the shapes
of the correlation functions by allowing the shapes to vary.
Therefore, it is not surprising that we find some degradation in the
$\Omega_M$ constraints with marginalization over the evolving bias
parameters.  On the other hand, all of the information about $m$ comes
from the relative amplitudes of \kapcor{} and \gamcor{} since $m$
simply scales these two functions relative to each other.
Consequently, allowing the shape of the correlation functions to vary
by marginalizing over the evolving bias parameters does not have a
significant impact on the $m$ constraints since the relative amplitude
information is preserved.

\citetalias{Jarvis:2015} used image simulations and a comparison of
two shear pipelines to constrain the multiplicative shear systematic
parameter to be $|m| \lesssim 0.05$.  The constraint on $m$ obtained
here is asymmetric around $m=0$ so it is somewhat difficult to compare
directly to the constraint obtained in \citetalias{Jarvis:2015}.  The
lower limit of our 68\% posterior interval in the evolving bias
analysis ($-0.05$) is roughly that obtained in
\citetalias{Jarvis:2015}.  However, the upper limit of our 68\%
confidence interval (0.80) is considerably weaker than that obtained
in \citetalias{Jarvis:2015}.  The width of our constraint on $m$ is
roughly a factor of 8 larger than that of \citetalias{Jarvis:2015}.
Note, though, that the constraint on $m$ presented here relies only on
the data, in contrast to the simulation-based approach of
\citetalias{Jarvis:2015}.  We discuss prospects for improvement of the
constraint on $m$ in \S\ref{sec:discussion}.

The bottom panel of Fig.~\ref{fig:main_constraints} shows the
constraints on $\Delta_z$ --- the systematic error parameter
describing a shift in the photometrically measured tracer galaxy
$N(z)$ --- in the evolving bias analysis.  The dashed (blue) curve
shows the constraint on $\Delta_z$ for the constant bias analysis.
The 68\% posterior interval on $\Delta_z$ in the evolving bias
analysis is $(-0.17, 0.07)$ while it is $(-0.38,0.04)$ for the
constant bias analysis.  Both of these intervals contain $\Delta_z =
0$, and are therefore consistent with no systematic bias of the tracer
galaxy $N(z)$.  Our constraints on $\Delta_z$ are consistent with the
analysis of \citet{Bonnett:2015}, which compared \texttt{skynet2}
photo-$z$ estimates to spectroscopically measured redshifts.
\citet{Bonnett:2015} found that difference between the mean photo-$z$
and the mean spectroscopic redshift was less than $\sim0.04$ for
galaxies in the DES SV `Gold' catalog.

It is a bit surprising that marginalization over the evolving bias
parameters causes the constraint on $\Delta_z$ to tighten slightly.
The explanation for this behavior can be deduced from
Fig.~\ref{fig:main_constraints}.  The posterior on $\Delta_z$ in the
constant bias model case exhibits two peaks, the larger centered at
$\Delta_z \approx 0$ and the smaller at $\Delta_z \approx -0.4$.  The two
peaks in the posterior result from weak tension between the \gamcor{}
and \kapcor{} measurements in the constant bias model for the fiducial
cosmological parameters.  Marginalization over the evolving bias
parameters allows the shape of \kapcor{} and \gamcor{} to vary
somewhat, relieving this tension and causing the posterior to prefer
$\Delta_z \approx 0$.  So by effectively excluding the peak at
$\Delta_z \approx -0.4$, the posterior that marginalizes over $a_i$ and
$c_i$ has lower variance than the posterior that keeps $a_i = c_i =
0$.  We emphasize that the tension between \kapcor{} and \gamcor{} is
weak and that it is only relevant to the constant bias analysis (which
is not expected to be a very good match to the data anyway).

Although our constraint on $\Delta_z$ is weaker than those of
\citet{Bonnett:2015} by a factor of a few, it was obtained without the
need for any spectroscopic calibration sample and used only $\sim3\%$
of the full DES survey area.  In many ways, this is one of the most
exciting results of this work: we have shown the photo-$z$
distribution can be tightly constrained using the joint measurement of
\kapcor{} and \gamcor{}.  The measurement presented here serves as a
test case for future applications of this potentially powerful
technique for estimating galaxy redshift distributions.

\begin{table*}
  \centering
  \caption{Priors and posteriors on the model parameters introduced in
    \S\ref{subsubsec:cosmo_param}, \S\ref{subsubsec:shear_bias_param},
    and \S\ref{subsubsec:redshift_bias_param}.  The three different
    analyses ({\it cosmology}, {\it shear calibration}, and {\it
      tracer redshift}) are each aimed at constraining a different
    parameter ($\Omega_M$, $m$, and $\Delta_z$); the posteriors on
    these parameters are shown in the last two columns.  The {\it
      constant bias} column corresponds to letting the bias parameter
    $b_0$ vary, but fixing $a_i=c_i=0$; the {\it evolving bias}
    column corresponds to letting $b_0$, $a_i$ and $c_i$ vary
    simultaneously.  All cosmological parameters not shown in the
    table are fixed to their best fit values from the CMB-only
    analysis of \citetalias{PlanckXIII:2015}.}
  \label{tab:params}
  \begin{tabular}{cccccc}
    Analysis Name & $\Omega_M$ & $\Delta_{z}$ & $m$ & 68\% posterior interval, constant bias & 68\% posterior interval, evolving bias\\
    \hline 
    Cosmology & $[0.05, 0.6]$ & 0.0 & 0.0 & $\Omega_M \in (0.25,0.33)$ & $\Omega_M \in (0.26, 0.38)$ \\
    Shear calibration& 0.3121 & 0.0 & [-3.0, 4.0] & $m \in (-0.07, 0.68)$ & $m \in (-0.05, 0.80)$ \\
    Tracer redshift & 0.3121 & [-0.5, 0.5] & 0.0 & $\Delta_z \in (-0.38,0.04)$ & $\Delta_z \in (-0.17,0.07)$ \\
  \end{tabular}
\end{table*}

\subsection{Effects of bias parameter marginalization}
\label{subsec:bias_marg}

The results shown as the solid (red) curves in
Fig.~\ref{fig:main_constraints} were obtained from model fits that
allowed the bias parameters $a_i$ and $c_i$ to vary, while the dashed
(blue) curves show the posteriors on our model parameters for the
constant bias analysis, i.e. when we fix $a_i = c_i = 0$.  Since the
$a_i$ affect the model prediction at small angular scales,
marginalization over these parameters effectively down-weights the
contribution of small scales to the likelihood.  It is clear from
Fig.~\ref{fig:main_constraints} that the evolving bias
marginalization has a fairly small effect on our results.

%Interestingly, the marginalization over $a_i$ and $c_i$ slightly
%reduces the width of the posterior on $\Delta_z$.  This is somewhat
%surprising as one would naively expect marginalization over additional
%parameters to increase the width of the posterior. The tail of the
%posterior for $\Delta_z$ in the $a_i = c_i = 0$ model is being driven
%by a region of moderate likelihood centered around $\Delta_z \sim
%-0.35$ that results from slight tension between the \kapcor{} and
%\gamcor{} measurements at small angular scales in the baseline
%(i.e. no systematics) $a_i = c_i = 0$ model.  When we marginalize over
%$a_i$ and $c_i$, this region of parameter space becomes disfavored
%relative to the new maximum likelihood point and the error bar on
%$\Delta_z$ decreases slightly. \EB{explain tension or get rid of this
%  statement}
 
\begin{figure}
  \includegraphics[width=1.02\columnwidth]{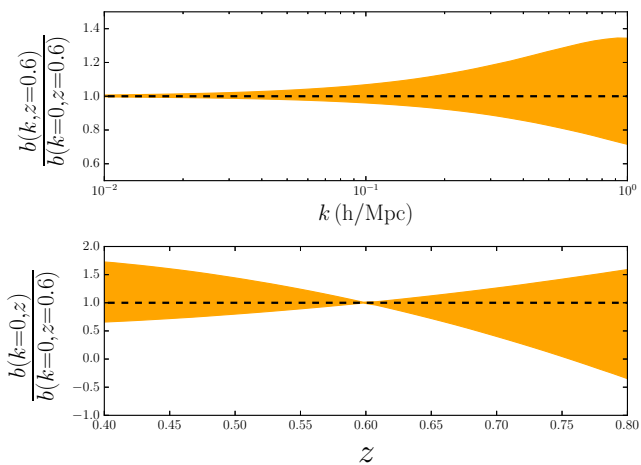}
  \caption{68\% confidence bands on the bias model $b(k,z)$
    (Eq.~\ref{eq:bias_model}) resulting from the cosmology analysis
    discussed in \S\ref{subsubsec:cosmo_param}.  In this analysis, we vary
    $\Omega_M$ and the bias parameters $b_0$, $a_i$ and $c_i$.  The
    bias models shown are normalized by $b(k=0,z=0.6)$ for
    clarity.  The data are consistent with no evolution of the bias in
    $k$ and $z$.}
  \label{fig:bias_constraints}
\end{figure}

Rather than marginalizing over $b_0$, $a_i$, and $c_i$, it is also
possible to consider the constraints on these parameters from the
analysis of \kapcor{} and \gamcor{}.  Constraints on these three bias
parameters translate into constraints on $b(k,z)$, and these
constraints are presented in Fig.~\ref{fig:bias_constraints}.  To
generate this figure, we drew $b_0$, $a_i$ and $c_i$ from the Markov
Chains of the cosmology analysis (the results of the other analyses
look similar).  Using these parameter values, we then evaluated the
resultant $b(k,z)$ across a range of $k$ (top panel) and $z$ (bottom
panel) values and the 68\% confidence band on $b(k,z)$ was determined.
We have normalized the results in both panels of
Fig.~\ref{fig:bias_constraints} by the large-scale (i.e. $k=0$) bias
at $z=0.6$ (roughly the center of the redshift distribution for the
tracer galaxies).  Fig.~\ref{fig:bias_constraints} shows that at 68\%
confidence, there is no evidence for departures from constant
bias (i.e. $a_i = c_i = 0$).

\section{Discussion}
\label{sec:discussion}

We have presented a joint measurement of galaxy-galaxy lensing and
galaxy-CMB lensing using DES SV imaging data and CMB lensing data from
the SPT-SZ survey.  The measurements of \kapcor{} and \gamcor{}
presented here are in agreement with other recent analyses of DES SV
data \citep{Giannantonio:2016, Crocce:2016, Clampitt:2016}.  We have
performed fits to the joint measurement of \kapcor{} and \gamcor{} to
extract constraints on cosmology and the presence of systematics in
the data.  In general, these fits illustrate that data from SPT and
DES are in good agreement with each other in the framework of the
currently favored flat-$\Lambda$CDM cosmological model from
\citetalias{PlanckXIII:2015}.

Assuming cosmology is well-constrained by other datasets, we have
shown that the joint measurement of \kapcor{} and \gamcor{} can be
used to constrain shear calibration as well as galaxy redshift
distributions, $N(z)$.  The constraints obtained on shear calibration
in this analysis have the advantage that they do not rely on
simulations of the shear measurement process.  The constraints
obtained on $N(z)$ have the advantage that they do not rely on
spectroscopic redshift measurements.  Encouragingly, our analysis
shows no strong evidence for systematics in either DES shear or
photo-$z$ measurements.

Additional data from DES will significantly improve the constraining
power of the joint lensing measurement presented here.  This analysis
used only data from the DES Science Verification region, a small
fraction (roughly 3\%) of the full survey area.  With additional DES
imaging, the region of useful overlap between DES and SPT will expand
significantly.  Ultimately, the overlap between the two surveys is
expected to be $\sim 2500$ sq. deg., roughly $19$ times larger than
the DES SV area.  Approximately, then, the constraints on cosmology
and systematics parameters obtained in this work can be expected to
tighten by a factor of $\sqrt{19} \sim 4.4$.  The improvement is
likely to be better than a factor 4.4 since the larger area of SPT and
DES overlap will also mean that the measurement of \kapcor{} and
\gamcor{} can be extended to larger angular scales. In this work we
set $\theta_{\rm max} = 50'$ to ensure that the maximum angular scale
probed was comparable to the size of the jackknife subregions.  With
larger area, the size of the jackknife subregions can be increased,
allowing $\theta_{\rm max}$ to be increased as well.  Since there is
additional signal at $\theta > 50'$, increasing $\theta_{\rm max}$
will improve parameter constraints.  Additionally, using full sky CMB
$\kappa$ maps from Planck \citep{Planck2015:XV}, it will be possible
to exploit the full 5000 sq. deg. area of the full DES survey area.
This represents a factor of two increase over the SPT and DES overlap,
albeit at somewhat lower signal to noise. Below, however, we take a
conservative approach and assume only a factor of 4.4 improvement in
the signal-to-noise with the final DES data set.

For the cosmology analysis, a factor of 4.4 improvement in the
signal-to-noise would result in $\delta \Omega_M \sim 0.009$,
comparable to the current error bar on $\Omega_M$ from the CMB-only
analysis of \citetalias{PlanckXIII:2015}.  Note, though, that this is
not really a fair comparison since our constraint on $\Omega_M$ was
derived after marginalizing over only the galaxy bias parameters,
while the \citetalias{PlanckXIII:2015} constraint marginalizes over
all the other $\Lambda$CDM parameters.  Still, there is reason to be
optimistic: as shown in \citet{Das:2009}, the ratio
$C_{\ell}^{\kappa_{{\rm Gal}} g}/C_{\ell}^{\kappa_{{\rm CMB}} g}$ does
not depend on the galaxy power specrum.  While the joint \kapcor{} and
\gamcor{} observable considered here does not exactly share this
property, we expect the joint lensing observable to be fairly robust
to cosmological parameters that change the shape of the galaxy power
spectrum, but that do not change the purely geometrical factor
$g^{{\rm CMB}}(\chi^g)/g^s(\chi^g)$.  A full cosmological analysis is
needed to quantify exactly how much marginalization over additional
cosmological parameters will degrade our constraint on $\Omega_M$ and
is beyond the scope of this work.  We also note that the constraints
on cosmology could be improved further by dividing the tracer and
source galaxies into more redshift bins.

A factor of $4.4$ tightening of our constraint on the multiplicative
shear bias, $m$, would yield $\delta m \sim 0.1$.  This is a factor of
two larger than the current constraints of $|m| \lesssim 0.05$ from
the analysis presented in \citetalias{Jarvis:2015}.  Given this
result, it is likely that shear calibration will continue to be
performed using image simulations.  That said, data-only analyses like
that considered here have a valuable roll to play as consistency
checks on such simulations.  Futhermore, future CMB data (see below)
could significantly improve the constraints on $m$ obtained with a
joint \kapcor{} and \gamcor{} analysis, making these constraints
competitive with simulation calibration.

Tightening our constraint on $\Delta_z$ by a factor of 4.4 would
result in $\delta (\Delta_z) \sim 0.02$.  This is roughly a factor of
two tighter than the constraint on $\Delta_z$ presented in
\citet{Bonnett:2015}.  Joint measurement of \kapcor{} and \gamcor{}
has the potential to be a highly competitive probe of photo-$z$
systematics.  Note that our $\Delta_z$ analysis was optimistic in the
sense that we did not vary $m$ and $\Delta_z$ simultaneously, and
these two systematics parameters are expected to be somewhat
degenerate.  However, given tight priors on $m$ from image
simulations, the constraints on $\Delta_z$ will not be significantly
degraded by marginalizing over $m$.

Future CMB lensing maps from SPT-3G \citep{Benson:2014} and Advanced
ACTPol \citep{Calabrese:2014} will significantly improve the
signal-to-noise of the CMB lensing measurements.  Since the CMB
lensing map used here is noise dominated at all but the largest
angular scales, future CMB lensing maps obtained with these
experiments will improve the signal-to-noise of the joint \kapcor{}
and \gamcor{} measurement beyond the factor of 4.4 considered above.
Such future measurements will be able to place interesting constraints
on cosmology as well as provide independent checks on the presence of
systematic errors in the data using the joint measurement of \kapcor{}
and \gamcor{}.

\section*{Acknowledgements}

This paper has gone through internal review by the DES collaboration.

Funding for the DES Projects has been provided by the U.S. Department
of Energy, the U.S. National Science Foundation, the Ministry of
Science and Education of Spain, the Science and Technology Facilities
Council of the United Kingdom, the Higher Education Funding Council
for England, the National Center for Supercomputing Applications at
the University of Illinois at Urbana-Champaign, the Kavli Institute of
Cosmological Physics at the University of Chicago, the Center for
Cosmology and Astro-Particle Physics at the Ohio State University, the
Mitchell Institute for Fundamental Physics and Astronomy at Texas A\&M
University, Financiadora de Estudos e Projetos, Funda{\c c}{\~a}o
Carlos Chagas Filho de Amparo {\`a} Pesquisa do Estado do Rio de
Janeiro, Conselho Nacional de Desenvolvimento Cient{\'i}fico e
Tecnol{\'o}gico and the Minist{\'e}rio da Ci{\^e}ncia, Tecnologia e
Inova{\c c}{\~a}o, the Deutsche Forschungsgemeinschaft and the
Collaborating Institutions in the Dark Energy Survey.

The Collaborating Institutions are Argonne National Laboratory, the
University of California at Santa Cruz, the University of Cambridge,
Centro de Investigaciones Energ{\'e}ticas, Medioambientales y
Tecnol{\'o}gicas-Madrid, the University of Chicago, University College
London, the DES-Brazil Consortium, the University of Edinburgh, the
Eidgen{\"o}ssische Technische Hochschule (ETH) Z{\"u}rich, Fermi
National Accelerator Laboratory, the University of Illinois at
Urbana-Champaign, the Institut de Ci{\`e}ncies de l'Espai (IEEC/CSIC),
the Institut de F{\'i}sica d'Altes Energies, Lawrence Berkeley
National Laboratory, the Ludwig-Maximilians Universit{\"a}t
M{\"u}nchen and the associated Excellence Cluster Universe, the
University of Michigan, the National Optical Astronomy Observatory,
the University of Nottingham, The Ohio State University, the
University of Pennsylvania, the University of Portsmouth, SLAC
National Accelerator Laboratory, Stanford University, the University
of Sussex, Texas A\&M University, and the OzDES Membership Consortium.

The DES data management system is supported by the National Science
Foundation under Grant Number AST-1138766.  The DES participants from
Spanish institutions are partially supported by MINECO under grants
AYA2012-39559, ESP2013-48274, FPA2013-47986, and Centro de Excelencia
Severo Ochoa SEV-2012-0234.  Research leading to these results has
received funding from the European Research Council under the European
Union's Seventh Framework Programme (FP7/2007-2013) including ERC
grant agreements 240672, 291329, and 306478.

We are grateful for the extraordinary contributions of our CTIO
colleagues and the DECam Construction, Commissioning and Science
Verification teams in achieving the excellent instrument and telescope
conditions that have made this work possible.  The success of this
project also relies critically on the expertise and dedication of the
DES Data Management group.

The South Pole Telescope program is supported by the National
Science Foundation through grant PLR-1248097. Partial support
is also provided by the NSF Physics Frontier Center grant
PHY-0114422 to the Kavli Institute of Cosmological Physics at the
University of Chicago, the Kavli Foundation, and the Gordon and
Betty Moore Foundation through Grant GBMF\#947 to the University
of Chicago.

Argonne National Laboratory's work was supported under the
U.S. Department of Energy contract DE-AC02-06CH11357.

\bibliographystyle{mnras}
\bibliography{thebibliography}

\section*{Affiliations}

\textit{
%\scriptsize
$^{1}$ Department of Physics and Astronomy, University of Pennsylvania, Philadelphia, PA 19104, USA\\
$^{2}$ Institute of Astronomy, University of Cambridge, Madingley Road, Cambridge CB3 0HA, UK\\
$^{3}$ Kavli Institute for Cosmology, University of Cambridge, Madingley Road, Cambridge CB3 0HA, UK\\
$^{4}$ Centre for Theoretical Cosmology, DAMTP, University of Cambridge, Wilberforce Road, Cambridge CB3 0WA, UK\\
$^{5}$ Fermi National Accelerator Laboratory, P. O. Box 500, Batavia, IL 60510, USA\\
$^{6}$ Kavli Institute for Cosmological Physics, University of Chicago, Chicago, IL 60637, USA\\
$^{7}$ Department of Physics, University of Michigan, Ann Arbor, MI 48109, USA\\
$^{8}$ Department of Physics, University of Chicago, 5640 South Ellis Avenue, Chicago, IL 60637, USA\\
$^{9}$ Argonne National Laboratory, Argonne, IL, 60439, USA\\
$^{10}$ Department of Astronomy and Astrophysics, University of Chicago, Chicago, IL 60637, USA\\
$^{11}$ Institut de Ci\`encies de l'Espai, IEEC-CSIC, Campus UAB, Carrer de Can Magrans, s/n,  08193 Bellaterra, Barcelona, Spain\\
$^{12}$ Department of Physics \& Astronomy, University College London, Gower Street, London, WC1E 6BT, UK\\
$^{13}$ Institut de F\'{\i}sica d'Altes Energies (IFAE), The Barcelona Institute of Science and Technology, Campus UAB, 08193 Bellaterra (Barcelona) Spain\\
$^{14}$ Department of Physics, Stanford University, 382 Via Pueblo Mall, Stanford, CA 94305, USA\\
$^{15}$ Kavli Institute for Particle Astrophysics \& Cosmology, P. O. Box 2450, Stanford University, Stanford, CA 94305, USA\\
$^{16}$ Jodrell Bank Center for Astrophysics, School of Physics and Astronomy, University of Manchester, Oxford Road, Manchester, M13 9PL, UK\\
$^{17}$ Cerro Tololo Inter-American Observatory, National Optical Astronomy Observatory, Casilla 603, La Serena, Chile\\
$^{18}$ Department of Physics and Electronics, Rhodes University, PO Box 94, Grahamstown, 6140, South Africa\\
$^{19}$ Department of Astrophysical Sciences, Princeton University, Peyton Hall, Princeton, NJ 08544, USA\\
$^{20}$ CNRS, UMR 7095, Institut d'Astrophysique de Paris, F-75014, Paris, France\\
$^{21}$ Sorbonne Universit\'es, UPMC Univ Paris 06, UMR 7095, Institut d'Astrophysique de Paris, F-75014, Paris, France\\
$^{22}$ Carnegie Observatories, 813 Santa Barbara St., Pasadena, CA 91101, USA\\
$^{23}$ Laborat\'orio Interinstitucional de e-Astronomia - LIneA, Rua Gal. Jos\'e Cristino 77, Rio de Janeiro, RJ - 20921-400, Brazil\\
$^{24}$ Observat\'orio Nacional, Rua Gal. Jos\'e Cristino 77, Rio de Janeiro, RJ - 20921-400, Brazil\\
$^{25}$ Department of Astronomy, University of Illinois, 1002 W. Green Street, Urbana, IL 61801, USA\\
$^{26}$ National Center for Supercomputing Applications, 1205 West Clark St., Urbana, IL 61801, USA\\
$^{27}$ Department of Physics, McGill University, 3600 rue University, Montreal, QC, H3A 2T8, Canada\\
$^{28}$ Excellence Cluster Universe, Boltzmannstr.\ 2, 85748 Garching, Germany\\
$^{29}$ Faculty of Physics, Ludwig-Maximilians-Universit\"at, Scheinerstr. 1, 81679 Munich, Germany\\
$^{30}$ Department of Astronomy, University of Michigan, Ann Arbor, MI 48109, USA\\
$^{31}$ SLAC National Accelerator Laboratory, Menlo Park, CA 94025, USA\\
$^{32}$ Department of Physics, University of California, Berkeley, CA, 94720, USA\\
$^{33}$ Center for Cosmology and Astro-Particle Physics, The Ohio State University, Columbus, OH 43210, USA\\
$^{34}$ Department of Physics, The Ohio State University, Columbus, OH 43210, USA\\
$^{35}$ Australian Astronomical Observatory, North Ryde, NSW 2113, Australia\\
$^{36}$ Departamento de F\'{\i}sica Matem\'atica,  Instituto de F\'{\i}sica, Universidade de S\~ao Paulo,  CP 66318, CEP 05314-970, S\~ao Paulo, SP,  Brazil\\
$^{37}$ George P. and Cynthia Woods Mitchell Institute for Fundamental Physics and Astronomy, and Department of Physics and Astronomy, Texas A\&M University, College Station, TX 77843,  USA\\
$^{38}$ Department of Astronomy, The Ohio State University, Columbus, OH 43210, USA\\
$^{39}$ Instituci\'o Catalana de Recerca i Estudis Avan\c{c}ats, E-08010 Barcelona, Spain\\
$^{40}$ Max Planck Institute for Extraterrestrial Physics, Giessenbachstrasse, 85748 Garching, Germany\\
$^{41}$ Jet Propulsion Laboratory, California Institute of Technology, 4800 Oak Grove Dr., Pasadena, CA 91109, USA\\
$^{42}$ School of Physics, University of Melbourne, Parkville, VIC 3010, Australia\\
$^{43}$ Department of Physics and Astronomy, Pevensey Building, University of Sussex, Brighton, BN1 9QH, UK\\
$^{44}$ Centro de Investigaciones Energ\'eticas, Medioambientales y Tecnol\'ogicas (CIEMAT), Madrid, Spain\\
$^{45}$ Brookhaven National Laboratory, Bldg 510, Upton, NY 11973, USA\\
$^{46}$ Harvard Smithsonian Center for Astrophysics, 60 Garden St, MS 12, Cambridge, MA, 02138, USA\\
$^{47}$ Institute of Cosmology \& Gravitation, University of Portsmouth, Portsmouth, PO1 3FX, UK\\
}

% Don't change these lines
\bsp	% typesetting comment
\label{lastpage}

\end{document}